\documentclass[hidelinks,onefignum,onetabnum]{siamart220329}

\usepackage[ruled,algo2e]{algorithm2e}
\usepackage{lipsum}
\usepackage{amsfonts}
\usepackage{epstopdf}
\usepackage{amsopn}

\DeclareMathOperator*{\argmin}{arg\,min}

\newcommand{\Rd}{\mathbb{R}^d}
\newcommand{\R}{\mathbb{R}}

\newcommand{\Vm}{V_m}

\ifpdf
  \DeclareGraphicsExtensions{.eps,.pdf,.png,.jpg}
\else
  \DeclareGraphicsExtensions{.eps}
\fi

\newsiamremark{remark}{Remark}
\newsiamremark{hypothesis}{Hypothesis}
\crefname{hypothesis}{Hypothesis}{Hypotheses}
\newsiamthm{claim}{Claim}

\headers{DL for Committor Functions with Adaptive Sampling}{B. Lin and W. Ren}

\title{Deep Learning Method for Computing Committor Functions with Adaptive Sampling}

\author{Bo Lin\thanks{Department of Mathematics, National University of Singapore, 10 Lower Kent Ridge Road, Singapore 119076 (\email{matboln@nus.edu.sg}, \email{matrw@nus.edu.sg}).}
\and Weiqing, Ren\footnotemark[1]}

\begin{document}

\maketitle

\begin{abstract}
The committor function is a central object for quantifying the transitions between metastable states of dynamical systems. Recently, a number of computational methods based on deep neural networks have been developed for computing the high-dimensional committor function. The success of the methods relies on sampling adequate data for the transition, which still is a challenging task for complex systems at low temperatures. In this work, we propose a deep learning method with two novel adaptive sampling schemes (I and II). In the two schemes, the data are generated actively with a modified potential where the bias potential is constructed from the learned committor function. We theoretically demonstrate the advantages of the sampling schemes and show that the data in sampling scheme II are uniformly distributed along the transition tube. This makes a promising method for studying the transition of complex systems. The efficiency of the method is illustrated in high-dimensional systems including the alanine dipeptide and a solvated dimer system.

\end{abstract}

% REQUIRED

% REQUIRED
% \begin{MSCcodes}
% ??
% \end{MSCcodes}

\section{Introduction}
Understanding the transition between metastable states of dynamical systems is of great importance in the field of computational science. 
Typical systems can be arising in bio-molecules, chemical reactions and nucleation process in phase transitions. 
The main objective in studying the noise-induced transitions or rare events is the mechanism underlying the transition, such as the transition state, transition rate and reactive pathway. 
However, predicting the transition mechanism for practical systems is usually a challenging task, due to difficulties including the complexity of the system with few knowledge regarding the transition, high dimensionality of the configuration space under consideration and long-waiting times for observing transition events in machine simulations. 
Over the past decades, a large number of methods have been developed for the study of rare events, with notable ones including 
% include notably Existing notable computational methods for investigating the transition mechanism include 
the nudged elastic band method~\cite{Newell81}, the string method~\cite{weinan2002string,ren2005transition,ren2007simplified} and transition path sampling~\cite{bolhuis2002transition,dellago2002transition}.

The transition path theory (TPT) provides a framework for quantitatively describing the transition events~\cite{vanden2006towards,vanden2010transition}, in which the committor function plays a central role~\cite{weinan2005transition}. 
For example, the transition states are located near the 1/2-isosurface of the committor function. 
% The committor function is a function defined in the configuration space which maps the states to a probability 
The committor function is a probability function defined in the configuration space, which is in high dimensions for practical systems such as molecular systems. To compute the committor function, a typical starting point is the partial differential equation (PDE) description of the function and its equivalent variational formulation~\cite{gardinerhandbook}. However, the high dimensionality of the system renders the finite difference and finite element methods inapplicable for solving this PDE, since the computational complexity and cost are prohibitively expensive. 
To make the computation of the committor function tractable, several methods have been proposed including the finite-temperature string method~\cite{ren2005transition}, the diffusion map method~\cite{lai2018point,trstanova2020local,evans2023computing} and the method using tensor networks~\cite{chen2023committor}.

Recently, a number of computational methods based on deep learning have been developed for computing the high-dimensional committor function~\cite{khoo2019solving,li2019computing,rotskoff2022active,hasyim2022supervised}. 
In the methods, the committor function is parameterized by a neural network.
% The ability of neural networks for representing high-dimensional functions enables the methods to deal with high-dimensional systems. 
The loss function was built as an unbiased estimator for the variational functional characterizing the committor function, where the data are generated by importance sampling from a known distribution. 
The success of the methods relies on sampling adequate data that covers the transition tube in which the transition takes place with high probability. 
To this end, several sampling methods have been proposed~\cite{li2019computing,rotskoff2022active,hasyim2022supervised}. For example, it was proposed to sample data from the system with raised temperature or using metadynamics~\cite{li2019computing}. 
Also, one can generate data using the adaptive sampling by making use of the learned committor function~\cite{rotskoff2022active,hasyim2022supervised}. In fact, these methods have some limitations since they require prior knowledge regarding the transition or may poorly explore the entire transition tube at low temperatures~\cite{hasyim2022supervised}. Therefore, sampling the entire transition tube is still a challenging task for complex systems at low temperatures.
% In fact, 
% This is a challenging task for complex systems at low temperatures, although several promising sampling methods have been proposed~\cite{li2019computing,rotskoff2022active,hasyim2022supervised}. For instance, it was proposed to sample data from the system with raised temperature or using metadynamics~\cite{li2019computing}. 
% Also, the data can be generated using the adaptive sampling by making use of the learned committor function~\cite{rotskoff2022active,hasyim2022supervised}. These methods require prior knowledge regarding the transition or may poorly explore the entire transition tube at low temperatures.

In this work, we propose a deep learning method with two novel adaptive sampling schemes (I and II) for computing the committor function. The data are generated actively using the two schemes with known distributions. 
% The sampling schemes purely make use of the knowledge provided by the learned committor function. 
In sampling scheme I, we introduce a one-dimensional variable which is a function depending on the learned committor 
% whose isosurfaces are distributing more uniformly in the configuration space 
and sample data by performing metadynamics with this variable. 
In sampling scheme II, we generate data with a modified potential which incorporates the free energy associated with the learned committor function. 
% the equilibrium distribution where the potential is modified with the free energy associated with the learned committor function. 
We present theoretical evidences to show the advantages of the sampling schemes and prove that the data in sampling scheme II are uniform along the transition tube.  
% The method enables efficient exploration and sampling of the transition process based on the learned committor function where no extra knowledge regarding the transition is required, which 
% This makes the method a promising tool for studying the transition of complex systems.
Therefore, this method provides an efficient way for exploration and sampling of the transition tube, thus studying the transition of complex systems. 
We illustrate the efficiency of the proposed method in three high-dimensional benchmark problems including an extended Mueller system, the alanine dipeptide and a solvated dimer system.

The paper is organized as follows. In Section~\ref{previous}, we introduce the background and recent deep learning methods. In Section~\ref{method}, we propose a method with two adaptive sampling schemes and analyze the data distribution for the schemes. We present numerical results in Section~\ref{example}. In Section~\ref{conclusion}, we draw the conclusions.

\section{Background and Existing Methods}\label{previous}
Consider the system whose dynamics is modelled by the overdamped Langevin equation:
\begin{equation}\label{Lan}
    d x_t = -\nabla V(x_t) dt + \sqrt{2 k_B T} d W_t,\quad t>0
\end{equation}
where $V(x)$ is the potential function, $k_B$ is the Boltzmann constant, $T$ is the temperature and $W_t$ is a standard Brownian motion. For simplicity, we let $\epsilon=k_B T$. Denote by $\Omega$ the configuration space under consideration. The equilibrium distribution of the system is known as the Gibbs-Boltzmann density function $\rho(x)=Z^{-1}\exp(-V(x)/\epsilon)$ where $Z$ is a normalization constant. 

We consider the general case where the system has two metastable states $A$ and $B$ in $\Omega$, which correspond to the local minima of $V$. At low temperatures such that the thermal energy is much lower than the energy barrier separating the two metastable states, the system will remain in one state for exponentially long times before making transition to the other state. Our objective is to study the transition between metastable states.

According to the transition path theory~\cite{vanden2006towards,vanden2010transition}, a central object for understanding the transition mechanism is the committor function. The committor function $q(x)$ is defined as the probability that the system starting from $x$ arrives at the state $B$ first rather than $A$, {\it i.e.}
\begin{equation}\label{def_q}
    q(x)=\textbf{Prob}[\tau_A(x)> \tau_B(x)],\quad x\in\Omega
\end{equation}
where $\tau_A(x)$ and $\tau_B(x)$ are the first hitting times of $A$ and $B$, respectively,
\begin{equation}
\begin{aligned}
    \tau_A(x) = \inf\{t\geq 0 :x_t\in A,\ x_0=x\},\\
    \tau_B(x) = \inf\{t\geq 0 :x_t\in B,\ x_0=x\}.
\end{aligned}
\end{equation}

To compute the committor function, one may consider a  mathematical description of the committor function, which is given by the backward Kolmogorov equation~\cite{gardinerhandbook} with Dirichlet boundary conditions,
% over the states $A$ and $B$,
\begin{equation}\label{PDE}
\begin{cases}
-\nabla V(x)\cdot \nabla q(x) + \epsilon \Delta q(x) = 0,\quad x\in \Omega \setminus (A\cup B) \\
q(x)=0,\ x\in \partial A;\quad q(x)=1,\ x\in \partial B.
\end{cases}
\end{equation}
The variational formulation for the equation is given by the optimization problem,
\begin{equation}\label{var}
\begin{aligned}
&\argmin_{q} \int_{\Omega \setminus (A\cup B)}|\nabla q(x)|^2 \rho(x) dx\\
&\ \textbf{s.t.}\quad q(x)=0,\ x\in\partial A;\quad q(x)=1,\ x\in\partial B.
\end{aligned}
\end{equation}

To address the boundary conditions, it was proposed to represent $q$ using a composite form that automatically satisfies the conditions so that the variational formulation can be transformed into an unconstrained problem~\cite{li2019computing}. Another way for addressing boundary conditions is the penalty method which is able to deal with irregular boundaries. In this method, the objective problem is unconstrained where a penalty term corresponding to the boundary conditions is added,
% To address the boundary conditions,  one can 
%
\begin{equation}\label{prob}
\begin{aligned}
    \argmin_{q} \int_{\Omega\setminus(A\cup B)}|\nabla q(x)|^2 \rho(x) dx + \lambda
    \bigg(\int_{A} |q(x)|^2 \rho_A(x)dx  + \int_{B} |1-q(x)|^2 \rho_B(x)dx\bigg),
\end{aligned}
\end{equation}
where $\lambda$ is a parameter to balance the two terms in the functional and $\rho_A(x)$ and $\rho_B(x)$ are particular probability distributions over $A$ and $B$, respectively, which is used for generating samples to represent the penalty term.

% it was proposed to introduce a composite form for the committor function $q$ with mollified indicator functions that automatically satisfies the    conditions~\cite{li2019computing}. Then one can transform the variational formulation into an unconstrained optimization problem using the composite form. A more general way is the penalty method which is able to deal with irregular boundaries. In this method, one can first sample two data sets in $A$ and $B$, respectively, and then solve the following unconstrained problem where a penalty term is added,
% in which representative dataset for $A$ and $B$ are sampled. The unconstrained problem corresponding to Eq.~\eqref{var} using the penalty method is given by
% where $\lambda$ is a parameter to balance the two terms in the functional and $d\mu_A(x)$ and $d\mu_B(x)$ are particular probability measures over $A$ and $B$, respectively. 

Recently, a number of computational methods based on deep learning have been developed for computing the high-dimensional committor function~\cite{khoo2019solving,li2019computing,rotskoff2022active,hasyim2022supervised}. The central idea to solve the problem~\eqref{prob} where the committor function is parameterized by a neural network $q_{\theta}(x)$. To build the loss function, one can sample data from a probability distribution $\tilde{\rho}(x)$ and take the following loss by reweighting the samples,
\begin{equation*}\label{loss0}
    % \argmin_{\theta} \frac{1}{C}\sum_{i=1}^{N} |\nabla q_{\theta}(x_i)|^2 \frac{\rho(x_i)}{\tilde{\rho}(x_i)} + \lambda \left(\frac{1}{N_A}\sum_{i=1}^{N_A} |q_{\theta}(x^A_i)|^2 + \frac{1}{N_B}\sum_{i=1}^{N_B} |q_{\theta}(x^B_i)-1|^2\right),
     \argmin_{\theta} \frac{1}{\mathcal{C}}\sum_{i=1}^{N} |\nabla q_{\theta}(X_i)|^2 \frac{\rho(X_i)}{\tilde{\rho}(X_i)} + \lambda \left(\frac{1}{N_A}\sum_{i=1}^{N_A} |q_{\theta}(X^A_i)|^2 + \frac{1}{N_B}\sum_{i=1}^{N_B} |q_{\theta}(X^B_i)-1|^2\right),
\end{equation*}
where $\mathcal{C}=\sum_{i=1}^{N} \rho(x_i)/\tilde{\rho}(x_i)$, $\{X_i\}_{i=1}^{N}$ are sampled from $\tilde{\rho}$ and $\{X^A_i\}_{i=1}^{N_A}$, $\{X^B_i\}_{i=1}^{N_B}$ are data sets sampled in 
% $\tilde{\rho}$, $\rho_A$, $\rho_B$ in the sets $\Omega\setminus(A\cup B)$, 
$A$, $B$, respectively. Note that the above loss can be regarded as an unbiased estimator for the functional~\eqref{prob}.

In the deep learning methods, one needs to sample adequate data covering the transition tube in which the transition events take place with high probability. This is a nontrivial and challenging task for complex systems at low temperatures. To compute the committor function at a low temperature $\epsilon$, the previous method in Ref.~\cite{li2019computing} proposed to sample data from the equilibrium distribution with a raised temperature $\epsilon'>\epsilon$ or using metadynamics with suitable collective variables. For the former part, one can generate trajectories by simulating the dynamics~\eqref{Lan} at $\epsilon'$ and collect data points along the trajectories. 

Another way for sampling data is the adaptive sampling method~\cite{rotskoff2022active,hasyim2022supervised}, in which the data are generated actively using the learned committor function $q_{\theta}(x)$. 
Specifically, one can generate data using umbrella sampling with $q_{\theta}$. Let $L$ be the number of sampling windows. For each window, a data set is sampled from the distribution $\rho(x)p_l(x)$, $1\leq l\leq L$, where $p_l$ is a function constructed from $q_{\theta}$ so that the sampling is concentrated on the isosurface of $q_{\theta}$ with a target value $q_l\in [0,1]$. 
The method was proposed to enhance sampling in the transition state region, but may
poorly explore the transition tube at low temperatures~\cite{hasyim2022supervised}. Alleviating this issue requires a fine tuning of the parameters in umbrella sampling, which is not straightforward or computationally expensive for complex systems.

In this work, we will introduce two adaptive sampling schemes that are able to produce uniform data along the transition tube, thus improving the accuracy of the computed committor function.

\section{Method}\label{method}
We parameterize the committor function by a neural network $q_{\theta}(x)$ where a sigmoid function $\sigma(z)=1/(1+\exp(-x))$ is acting on the output layer of the network, {\it i.e.}
\begin{equation*}
    q_{\theta}(x) = \sigma(\tilde{q}_{\theta}(x)).
\end{equation*}
To impose the boundary conditions over the metastable states $A$ and $B$, we sample two data sets $\{X^A_i\}_{i=1}^{N_A}$, $\{X^B_i\}_{i=1}^{N_B}$ in $A$ and $B$, respectively, and 
employ the penalty losses
% in the loss which will be introduced in the following sections,
\begin{equation}\label{loss_bd}
\begin{aligned}
    L_A(\theta)&= \frac{1}{N_A}\sum_{i=1}^{N_A}\bigg[\Big|q_{\theta}(X_i^A)\Big|^2 + \text{ReLU}^2\left(\Tilde{q}_{\theta}(X_i^A)-\Tilde{q}_0\right)\bigg],\\
    L_B(\theta)&=\frac{1}{N_B}\sum_{i=1}^{N_B}\bigg[\Big|q_{\theta}(X_i^B)-1\Big|^2+\text{ReLU}^2\left(\Tilde{q}_1-\Tilde{q}_{\theta}(X_i^B)\right)\bigg],
\end{aligned}
\end{equation}
where $\Tilde{q}_0$, $\Tilde{q}_1$ are two scalar numbers such that $\sigma(\Tilde{q}_0)\approx 0$ and $\sigma(\Tilde{q}_1)\approx 1$. The two numbers represent the objective values for $\tilde{q}$ over $A$ and $B$, respectively. For instance, one can choose $\tilde{q}_0=-5$ and $\tilde{q}_1=5$. In the loss, the $\text{ReLU}^2$ term is added in order to deal with vanishing gradients which may arise in the first term ($|q_{\theta}(X_i^A)|^2$ or $|q_{\theta}(X_i^B)-1|^2$) of the loss.

We compute the committor function using the adaptive sampling~\cite{rotskoff2022active,hasyim2022supervised}, in an iterative procedure including the two steps:
\vspace{1.5mm}
\begin{itemize}
\item generating data using $q_{\theta}(x)$ by importance sampling;
\item further training $q_{\theta}(x)$ using a loss function where the samples are reweighted.
\end{itemize}
\vspace{1.5mm}
Next, we propose two sampling schemes (I and II) for the procedure and analyze the data distribution of the two schemes along the transition tube of the system.

\begin{figure}[t!]
\centering
\includegraphics[width=\linewidth]{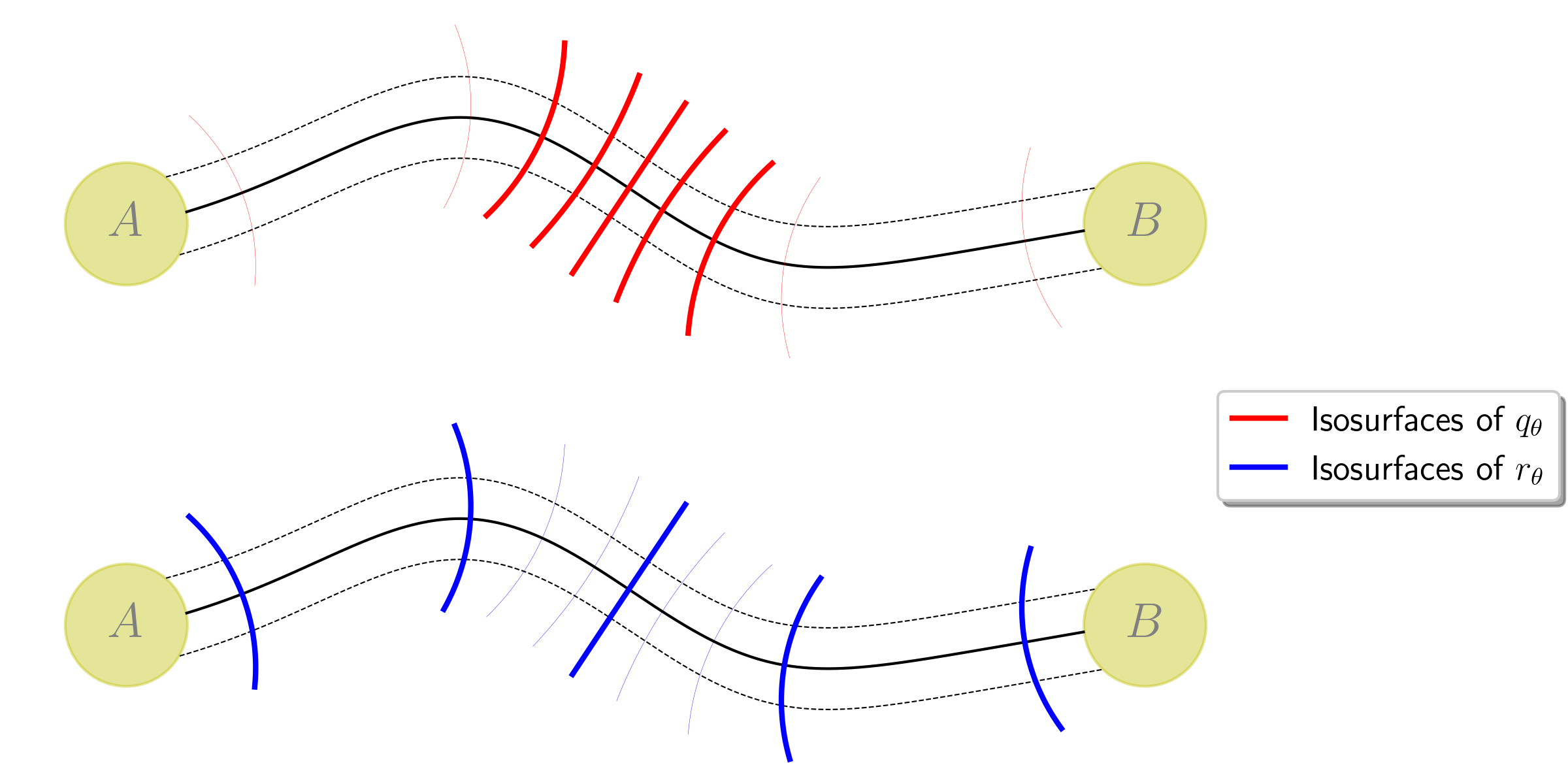}
\caption{Schematic illustration of the transition tube between metastable states $A$ and $B$, which is introduced in Section~\ref{Stability}, and the isosurfaces of the learned committor function $q_{\theta}(x)$ ({\bf Upper}), as well as of the variable $r_{\theta}(x)=R_5(q_{\theta}(x))$ ({\bf Lower}). In both panels, the isosurfaces represented by bold lines correspond to the function values $(2k-1)/10$, $1\leq k\leq 5$.}
\label{fig00}
\end{figure}

\subsection{Sampling Scheme I}\label{sec31}

With the learned committor function $q_{\theta}(x)$, we introduce a one-dimensional variable which is a function depending on $q_{\theta}$ for a given parameter $n\geq 1$,
\begin{equation}\label{Eq_r}
    r_{\theta}(x) = R_n(q_{\theta}(x)) := \frac{q_{\theta}(x)^{1/n}}{q_{\theta}(x)^{1/n}+(1-q_{\theta}(x))^{1/n}}.
\end{equation}
Note that the function $R_n(z)$ is an increasing function over $z\!\in\![0,1]$ with $R_n(0)\!=\!0$ and $R_n(1)\!=\!1$. In the special case where $n=1$, the variable $r_{\theta}(x)$ coincides with $q_{\theta}(x)$. A schematic illustration of the isosurfaces of $q_{\theta}(x)$ and $r_{\theta}(x)$ with $n=5$ in the configuration space is shown in \cref{fig00}. 
In the following, we will sample data using the metadynamics with the variable $r_{\theta}$.
% the one-dimensional variable $r_{\theta}(x)$ and sample data points with the modified potential energy surface.

For a given set of collective variables, the metadynamics is an iterative process of depositing Gaussian hills in the space of the variables to form a bias potential and exploring the free energy surface of the variables using the bias potential~\cite{laio2002escaping,barducci2011metadynamics}. It provides an efficient way for lowering the potential barrier separating the metastable states. Here, we perform metadynamics with the one-dimensional variable $r_{\theta}$. Specifically, the bias potential
% constructed from the deposited Gaussian hills 
at time $t$ is given by
\begin{equation}\label{meta}
\begin{aligned}
\Vm(x;t) &= G(r_{\theta}(x);t):= \sum_{t'=0,\tau,2\tau,\dots}^{t'<t} h\cdot \exp\left(-\frac{|r_{\theta}(x)-r_{\theta}(x_{t'})|^2}{2w^2}\right),
\end{aligned}
\end{equation}
where $h$, $w$ denote the height and width of the Gaussian hills and $\tau^{-1}$ indicates the frequency of the addition. For computational convenience, we discretize the range $[0,1]$ for $r_{\theta}$ using a fine mesh, and accumulate the values of the function $G(r;t)$ and its derivative $G'(r;t)$ over the mesh points. Then for a state $x$, the bias potential $V_m(x;t)$ and its gradient $\nabla V_m(x;t)$ can be approximated by the stored values $G(r_{*};t)$ and $G'(r_{*};t)\nabla r_{\theta}(x)$ where $r_{*}$ is the nearest mesh point to $r_{\theta}(x)$.

In metadynamics, the bias potential in Eq.~\eqref{meta} discourages the system to revisit the states explored formerly. After a sufficiently long time $\mathcal{T}$, the potential wells containing the metastable states are filled up by Gaussian hills, allowing the system to migrate from one state to another one frequently. 
For suitable hill height $h$, width $w$ and frequency $\tau^{-1}$, the bias potential in metadynamics provides a close estimate of the free energy $F_{r_{\theta}}(z)$ associated with the variable $r_{\theta}$ after a long time $\mathcal{T}$~\cite{barducci2011metadynamics},
\begin{equation}\label{approx}
    V_m(x;\mathcal{T}) \approx - F_{r_{\theta}}(r_{\theta}(x)) + E_0,
\end{equation} 
where $E_0$ is an additive constant and the free energy $F_{r_{\theta}}(z)$ is defined as
\begin{equation}
F_{r_{\theta}}(z)\!=\!-\epsilon\log\int_{\Omega}\rho(y)\delta(r_{\theta}(y)-z) dy.
\end{equation} 
In the following, we denote by $V_m(x)$ the bias potential at time $\mathcal{T}$ obtained from the metadynamics. 

We sample data points $\mathcal{D}=\{X_i\}_{i=1}^N$ from the equilibrium distribution over the region $\Omega\setminus(A\cup B)$,
\begin{equation}\label{rho_1}
    \rho_{\text{I}}(x)\sim\exp\left(-\frac{1}{\epsilon}\Big[V(x)+V_m(x)\Big]\right),
\end{equation}
with the modified potential $V(x)+V_m(x)$ in the system, where ``$\sim$'' denotes the proportional relationship between two functions. This can be done by generating trajectories following the dynamics with the modified potential and collect states along the trajectories. 

{\bf Remark:} {\it
It can be shown from the relation~\eqref{approx} that the sampling distribution $\rho_{\text{I}}(x)$ projected on $r_{\theta}(x)$ is uniform on the range $[0,1]$. In particular, in the case where $n=1$, the variable $r_{\theta}$ coincides with the learned committor function $q_{\theta}(x)$, in which performing metadynamics with $r_{\theta}$ is equivalent to the method of umbrella sampling with uniform target values of $q_{\theta}$.}

\begin{algorithm}[h!]
\caption{Computing the committor function using sampling scheme I.}
\label{alg1}
Initialize parameters $\theta=\theta^0$ in $q_{\theta}(x)$;\\
Choose function $R_n$, hill width $w$, height $h$ and frequency $\tau^{-1}$ for metadynamics;\\
% K$, $J$, $T$, $N$, $N_B$ and $\eta$.\\
\For{$k$ = $1$ to $K$}{
    Perform metadynamics with the variable $r_{\theta^{k-1}}(x)=R_n(q_{\theta^{k-1}}(x))$ until time $T$;\\
    Sample a dataset $\mathcal{D}$ from $\rho_{\text{I}}(x)$ in Eq.~\eqref{rho_1} with the modified potential;\\
    % by simulating the dynamics 
    % \hspace*{2mm}  modified potential $V(x)+V_m(x)$;\\
    Set $\theta^k_0=\theta^{k-1}$;\\
    \For{$i$ = $1$ to $I$}{
        Sample a batch $\mathcal{B}$ from $\mathcal{D}$;\\
        Compute $\nabla_{\theta}L(\theta^k_{i-1};\mathcal{B})$ as in Eq.~\eqref{loss1};\\
        Update $\theta^k_i=\theta^k_{i-1}-\eta\nabla_{\theta}L(\theta^k_{i-1};;\mathcal{B})$ or using Adam optimizer;
    }
    Set $\theta^k=\theta^k_I$.
}
{\bf Output}: The learned committor function $q_{\theta^K}(x)$.
\end{algorithm}

We carry out a further training of the neural network $q_{\theta}(x)$ using the loss function
\begin{equation}\label{loss1}
\begin{aligned}
      L(\theta;\mathcal{D}) = C^{-1}\sum_{i=1}^N |\nabla q_{\theta}(X_i)|^2 \exp\left(\frac{1}{\epsilon}\Vm(X_i)\right) + \lambda\left(
      L_A(\theta)+L_B(\theta)\right),
\end{aligned}
\end{equation}
where $C=\sum_{i=1}^N\exp\left(\Vm(X_i)/\epsilon\right)$, $\lambda$ is a parameter to balance the two terms in the loss and the penalty term is taken from Eq.~\eqref{loss_bd}. 
Note that the coefficients in the loss are accounted for the difference between the sampling and objective distributions, which makes the loss an unbiased estimator for the variational functional in Eq.~\eqref{prob}. 

In the method, we update the parameters in the neural network using the stochastic gradient descent. A pseudocode for the whole algorithm with $K$ iterations is presented in \cref{alg1}.

\subsection{Distribution of the Data in Scheme I}\label{Stability}
In sampling scheme I, the distribution of the data after convergence where $q_{\theta}(x)=q(x)$ is given by
\begin{equation}\label{p1}
	p_{\text{I}}(x)\sim \exp\left(-\frac{1}{\epsilon}\Big[V(x)-F_r(r(x))\Big]\right),
\end{equation}
which is obtained by replacing the bias potential $V_m$ in Eq.~\eqref{rho_1} by $-F_r$ due to the relation~\eqref{approx}, where $F_r$ is the free energy associated with the variable $r(x)=R_n(q(x))$ as in Eq.~\eqref{Eq_r}. Next, we analyze how the data are distributed along the transition tube of the system.

The transition tube is a connected region between $A$ and $B$ in the configuration space, which contains most of the reactive trajectories. It is defined via iso-surfaces of the committor function.
Denote by $\Gamma_z$ the isocommittor surface for a given committor value $z\in[0,1]$, {\it i.e.} $\Gamma_z=\{x\in \Omega:q(x)=z\}$.
The transition tube can be characterized by a set of probability density functions $\rho_z(x)$, $z\in[0,1]$, where $\rho_z(x)$ is the equilibrium distribution of the system restricted to the surface $\Gamma_z$. 
% {\color{red} width?} 
In particular, a centerline of the transition tube is defined as \begin{equation}\label{centerline}
	\begin{aligned}
		\varphi(z) = \int_{\Gamma_z} x \rho_z(x) dx,\quad 0\leq z\leq 1.
	\end{aligned}
\end{equation} 

The reparameterized centerline using its normalized arclength $\alpha$ is given by $\varphi(\alpha)$, $0\!\leq\! \alpha \!\leq \! 1$,  where $\varphi(0)\in A$, $\varphi(1)\in B$ and $|\varphi'(\alpha)|$ is constant along the curve. 
% One can compute an approximated committor function $\tilde{q}(x)$ 
One can obtain an approximated committor function $\tilde{q}(x)$ in the transition tube by solving the variational problem~\eqref{var} restricted to the tube and over a particular function space~\cite{ren2005transition}. 
The function space is a set of functions whose iso-surfaces are hyperplanes normal to the centerline. The function $\tilde{q}(x)$ restricted to the centerline, denoted by $f(\alpha)=\tilde{q}(\varphi(\alpha))$, is given by
\begin{equation}\label{f_alp}
	f(\alpha)=\frac{\int_{0}^{\alpha}\exp(F(\beta)/\epsilon) d\beta}{\int_{0}^1\exp(F(\beta)/\epsilon) d\beta},\quad \alpha\in[0,1]
\end{equation}
where the energy $F(\alpha)=-\epsilon\log\int_{\Omega} \rho(x) \delta((x-\varphi(\alpha))\cdot \vec{n})dx$ in which $\vec{n}$ denotes the tangent vector along the centerline $\varphi(\alpha)$. Indeed, the energy $F(\alpha)$ is the free energy associated with the tube variable $\tilde{s}(x)$ defined in the following.

% Before we proceed, we introduce the following variable which indicates the position for a given configuration $x$ when projected on the centerline by the iso-surface of $\tilde{q}(x)$.

\vspace{3mm}
{\bf Definition.} {\it The tube variable is defined as $\tilde{s}(x)=f^{-1}(\tilde{q}(x))\in[0,1]$ for a state $x$ in the transition tube, or $s(x)=f^{-1}(q(x))$, as the function $\tilde{q}(x)$ is close to original committor function $q(x)$.}
\vspace{3mm}

The tube variable $\tilde{s}(x)$ indicates the projected position for a given configuration $x$ in the transition tube, when $x$ is projected on the centerline through the hyperplanes of $\tilde{q}$. For each isocommittor surface $\Gamma_z$, the data distribution $p_{\text{I}}(x)$ in Eq.~\eqref{p1} restricted on the surface coincides with the equilibrium distribution $\rho_z(x)$, which indicates that the distribution $p_{\text{I}}(x)$ is concentrated in the transition tube. Furthermore, as shown in \cref{App_A}, the distribution projected on the tube variable $s(x)$ is given by
% $p(x)$ along the transition tube can be described by projecting it at the tube variable $s(x)$, which is given by

% can be written as 
% \begin{equation*}
% 	p(x) \sim c_z \exp\left(-\frac{1}{\epsilon}V(x)\right),\quad \text{for } x\in \Gamma_z
% \end{equation*}
% where $c_z=\exp\left(F_r(R_n(z))/\epsilon\right)$. This is exactly the equilibrium distribution $\rho_z(x)$ over the surface $\Gamma_z$, which implies the the sampling distribution $p(x)$ concentrates {\color{red}concentrates?} inside the transition tube.
% {\color{red} need to change notation X}
% Let $\mathfrak{X}$ be a random variable in the configuration space with the probability distribution $p(x)$. As the function $\tilde{s}(x)=f^{-1}(\tilde{q}(x))\in[0,1]$ gives the projected position on the centerline for a given configuration $x$, we view the distribution of $\tilde{s}(\mathfrak{X})$ as the projected sampling distribution for sampling scheme I on the centerline of the transition tube.

% Since $\tilde{q}(x)$ is close to the committor function $q(x)$, the distribution of $\tilde{s}(\mathfrak{X})$ can be estimated by the distribution of a variable $s(\mathfrak{X}):=f^{-1}(q(\mathfrak{X}))$, which is given by
\begin{equation}\label{g1}
g_{\text{I}}(z)\sim\exp(F(z)/\epsilon)R_n'(f(z)),\quad z\in[0,1].
\end{equation}
% \begin{equation}\label{distr}
%     g(s)\sim f'(s)R_n'(f(s))\sim\exp(F(s)/\epsilon)R_n'(f(s)),\quad s\in[0,1].
% \end{equation}
For a given function $R_n(\cdot)$, the density function $g_{\text{I}}(z)$ quantitatively measures how the data points generated from sampling scheme I are distributed along the transition tube. For a simple case where $n=1$, 
the function $R_n(\cdot)$ is identity. The density function $g_{\text{I}}(z)\!\sim\!\exp(F(z)/\epsilon)$ peaks in the transition state region where $F(z)$ attains its maximum value and becomes exponentially small otherwise when the temperature $\epsilon$ is low.
% especially in the two metastable sets $A$ and $B$,  
In this case, the data in sampling scheme I are inadequate for covering the entire transition tube. Note that this case also explains the inefficiency of the umbrella sampling with uniform target values of $q_{\theta}$ at low temperatures~\cite{rotskoff2022active,hasyim2022supervised}.
% is inefficient for exploring the entire transition tube. {\color{red} Note that this case also explains the inefficiency of umbrella sampling with uniform target values of $q_{\theta}$ at low temperatures.}

On the other hand, in the case where $n$ is greater than $1$, the function $R_n'(f(z))$ as defined in Eq.~\eqref{Eq_r} which achieves the minimum value at $f(z)=1/2$ or in the transition state region 
alleviates the disparities of the density function $g_{\text{I}}(z)$ along the transition tube. 
% Therefore, sampling scheme I with suitable parameter $n>1$ is capable of generating nearly uniform data along the transition tube. 
Therefore, choosing a suitable parameter $n>1$ in sampling scheme I can enhance the spreading out of sampling densities from the transition state region towards the entire transition tube (see \cref{fig00}).  
We will illustrate the distribution of the data in sampling scheme I in three numerical examples in Section~\ref{example}.

% Therefore, choosing a suitable parameter $n>1$ in sampling scheme I can enhance the 
% Therefore, in sampling scheme I, choosing a suitable number $n>1$ is able to enhance the sampling efficiency and thus improve the accuracy of the method. We demonstrate the effect of different choices for $n$ on the performance of the algorithm by numerical examples in Section~\ref{example}. 

\subsection{Sampling Scheme II and Distribution of the Data}
In this section, we introduce the second sampling scheme for the deep learning method. With a learned committor function $q_{\theta}(x)$, we first compute the free energy associated with $q_{\theta}$, denoted by $F_{q_{\theta}}(z)$, and then generate data points $\mathcal{D}=\{X_i\}_{i=1}^N$ from the equilibrium distribution 
\begin{equation}\label{rho_2}
    \rho_{\text{II}}(x)\sim \exp\left(-\frac{1}{\epsilon}\left[V(x)-\frac{1}{2}F_{q_{\theta}}(q_{\theta}(x))\right]\right),
\end{equation} 
with the modified potential $V(x)-\frac{1}{2}F_{q_{\theta}}(q_{\theta}(x))$ in the system. 
% Denote the coefficients $\xi_i=\exp\left[F_{q_{\theta}}(q_{\theta}(X_i))/(2\epsilon)\right]$, $1\leq i\leq N$.
We further update the parameters in the network using the loss function
\begin{equation}\label{loss2}
   J(\theta;\mathcal{D}) = D^{-1}\sum_{i=1}^N |\nabla q_{\theta}(X_i)|^2 \exp\left(-\frac{1}{2\epsilon}F_{q_{\theta}}(q_{\theta}(X_i))\right) + \lambda\left(
      L_A(\theta)+L_B(\theta)\right),
\end{equation}
where $D=\sum_{i=1}^N \exp\left[-F_{q_{\theta}}(q_{\theta}(X_i))/(2\epsilon)\right]$ and $\lambda$ is a parameter specifying the strength of the penalty term which is taken from Eq.~\eqref{loss_bd}. 
Note that we ignore the dependence of the coefficients  on the network parameters $\theta$ when computing gradient of the loss function. 
% Note that the loss~\eqref{loss2} provides an unbiased estimator for the variational functional in Eq.~\eqref{prob}.

% To analyze the distribution of the data sampled from the scheme~\eqref{rho_2}, consider the corresponding distribution with the original committor function $q(x)$,
% To analyze the distribution of the data along the transition tube, 
For the sampling scheme~\eqref{rho_2}, we consider the following distribution of the data after convergence where $q_{\theta}(x)=q(x)$,
\begin{equation}\label{p2}
    p_{\text{II}}(x) \sim \exp\left(-\frac{1}{\epsilon}\left[V(x)-\frac{1}{2} F_{q}(q(x))\right]\right),
\end{equation}
where $F_q$ is the free energy associated with $q$. Similarly, it can be verified that the distribution $p_{\text{II}}(x)$ is concentrated in the transition tube. Furthermore, as shown in \cref{App_A}, the distribution $p_{\text{II}}(x)$ projected on the tube variable $s(x)$ is approximately uniform, {\it i.e.}
\begin{equation}\label{g2}
    g_{\text{II}}(z)\approx 1,\quad z\in [0,1]
\end{equation}
which indicates that the sampling scheme is able to produce uniform data points along the transition tube.

% the distribution projected on the centerline can be represented by the distribution of a variable $s(\mathfrak{Y})$. 
%  Let $\mathfrak{Y}$ be a random variable in the configuration space with the distribution $p(x)$. 
% This demonstrates the efficiency of the sampling scheme for exploring the whole transition tube.
% $\mathfrak{Z}$

To obtain a close estimate of the free energy $F_{q_{\theta}}(z)$, we first compute the free energy $F_{r_{\theta}}(z)$ associated with the variable $r_{\theta}(x)=R_n(q_{\theta}(x))$ for a given number $n$ as in Eq.~\eqref{Eq_r}. This is done by metadynamics as introduced in Section~\ref{sec31}. Then the free energy $F_{q_{\theta}}(z)$ and its derivative can be represented as
\begin{equation}\label{r_q}
\begin{cases}
     F_{q_{\theta}}(z) = F_{r_{\theta}}(R_n(z))-\epsilon\log R_n'(z),\\
     F_{q_{\theta}}'(z) = F_{r_{\theta}}'(R_n(z))R_n'(z)-\epsilon\dfrac{R_n''(z)}{R_n'(z)}.
\end{cases}
\end{equation}
A pseudocode for the deep learning method with sampling scheme II is presented in \cref{alg2}.

\begin{algorithm}[h!]
\caption{Computing the committor function using sampling scheme II.}
\label{alg2}
Initialize parameters $\theta=\theta^0$ in $q_{\theta}(x)$;\\
Choose function $R_n$, hill width $w$, height $h$ and frequency $\tau^{-1}$ for metadynamics;\\
% Choose $K$, $J$, $T$, $N$, $N_B$ and $\eta$.\\
\For{$k$ = $1$ to $K$}{
    Perform metadynamics with variable $r_{\theta^{k-1}}(x)=R_n(q_{\theta^{k-1}}(x))$ until time $T$;\\
    Compute free energy $F_{r_{\theta}}(z)$ as in Eq.~\eqref{approx};\\
    Compute $F_{q_{\theta}}(z)$, $F_{q_{\theta}}'(z)$ using Eq.~\eqref{r_q};\\
    Sample a dataset $\mathcal{D}$ from $\rho_{\text{II}}(x)$ in Eq.~\eqref{rho_2} with the modified potential;\\
    % \\ \hspace*{2mm} the modified potential $V(x)-F_{q_{\theta}}(q_{\theta^{k-1}}(x))/2$;\\
    Set $\theta^k_0=\theta^{k-1}$;\\
    \For{$i$ = $1$ to $I$}{
        Sample a batch $\mathcal{B}$ from $\mathcal{D}$;\\
        Compute $\nabla_{\theta}J(\theta^k_{i-1};\mathcal{B})$ as in Eq.~\eqref{loss2};\\
        Update $\theta^k_i=\theta^k_{i-1}-\eta\nabla_{\theta}J(\theta^k_{i-1};\mathcal{B})$ or using Adam optimizer;
    }
    Set $\theta^k=\theta^k_I$.
}
{\bf Output}: The learned committor function $q_{\theta^K}(x)$.
\end{algorithm}

\section{Numerical Examples}\label{example}
To illustrate the efficiency of the method, we apply \cref{alg1} and \cref{alg2} to three systems including an extended Mueller system, the alanine dipeptide and a solvated dimer system. In the first example, we quantitatively evaluate the learned committor function with a reference solution which can be computed using the finite element method. In the second and third examples, the learned committor function is evaluated via a Monte-Carlo estimation for states sampled on particular isocommittor surfaces, especially on the transition state surface ($q=1/2$).

For the neural network architecture, in the first example, we use a fully-connected neural network with two hidden layers and $50$ nodes on each layer to parameterize the committor function. In the second and third examples, to enhance the training efficiency, we set the network so that it is equivalent to translation and rotation of the system in configuration space. Additionally, the network in the third example is chosen to be invariant to particle re-indexing due to the physical property of the system~\cite{schutt2017schnet}. In all of the examples, we take the activation function as the hyperbolic tangent ($\tanh$) and put a sigmoid function on the output layer to make sure the output is on $[0,1]$.

To impose the boundary conditions over the two metastable states $A$ and $B$, we sample two representative data sets $\mathcal{D}_A$ and $\mathcal{D}_B$, respectively. Then we initialize the neural network $q_{\theta}$ with $\mathcal{D}_A$, $\mathcal{D}_B$ by minimizing the loss function $L_A(\theta)+L_B(\theta)$ as defined in Eq.~\eqref{loss_bd} until $q_{\theta}$ closely satisfies the boundary conditions in $A$ and $B$,
% the loss is small, $L_A(\theta)+L_B(\theta)<10^{-4}$. 
\begin{equation*}
    E_{AB}(q_{\theta}) = \sqrt{\frac{1}{|\mathcal{D}_A|}\sum_{X\in\mathcal{D}_A}q_{\theta}(X)^2}+\sqrt{\frac{1}{|\mathcal{D}_B|}\sum_{X\in\mathcal{D}_B}(1-q_{\theta}(X))^2}<10^{-2}.
\end{equation*}
In the training process in \cref{alg1} and \cref{alg2}, we generate data using sampling scheme I or II and train the network for $5000$ steps using the stochastic gradient descent method with the Adam optimizer~\cite{kingma2014adam} and a mini-batch of $5000$ samples (in the first and second examples) or $100$ samples (in the third example).  The learning rate is set as $10^{-4}$. In the loss function as in Eq.~\eqref{loss1} and Eq.~\eqref{loss2}, we take the parameter $\lambda=1$. 
% The learning rate is set as $10^{-4}$ and the data set is split into a training set ($70\%$) and validation set ($30\%$). All of the codes are run in a machine with 3.20-GHz Intel Xeon W-3245 CPU and 11019-MB GeForce RTX 2080Ti graphics card.
% The training is stopped once the test error is above twice its minimum value and we select the neural network model with the lowest test error.

\subsection{Extended Mueller Potential}\label{Mueller}
In the first example, we consider an extended Muller system in the ten-dimensional space where the potential function is given by
% We first consider a ten-dimensional system with the extended Mueller potential composed of a rugged Mueller potential for the first two coordinates and harmonic ones for the remaining coordinates. The form of the extended Mueller potential is given by
\begin{equation}\label{example1}
\begin{aligned}
    V(x_1,\dots,x_{10}) &= V_{M}(x_1,x_2) + \frac{1}{2\sigma^2}\sum_{i=3}^{10}x_i^2,\\
    V_{M}(x_1,x_2) &= \sum_{i=1}^4 D_i\exp\Big[a_i(x_1-X_i)^2+b_i(x_1-X_i)(x_2-Y_i)\\
    &\qquad\qquad +c_i(x_2-Y_i)^2\Big]
     +\gamma \sin(2k\pi x_1)\sin(2k\pi x_2),
\end{aligned}
\end{equation}
which is a sum of the two-dimensional rugged Mueller potential $V_{M}(x_1,x_2)$ for the first two dimensions and the other eight harmonic ones for the remaining dimensions. The parameters $\gamma$, $k$ control the roughness of the potential function and $\sigma$ specifies the strength of the harmonic terms in the potential. In the example, we set $\gamma=9$, $k=5$, $\sigma=0.05$ and the other parameters in the potential function are taken from Ref.~\cite{li2019computing}.

We set the temperature as $\epsilon=10$ and define the two metastable sets $A$ and $B$ as 
\begin{equation}
\begin{aligned}
     A &= \{(x_1,\dots,x_{10})\in \R^{10}:|(x_1,x_2)- a|<r\},\\
     B &= \{(x_1,\dots,x_{10})\in \R^{10}:|(x_1,x_2)-b|<r\},
\end{aligned}
\end{equation}
where $a=(-0.558,1.441)$, $b=(0.623,0.028)$ and $r=0.1$. The two points $a$, $b$ correspond to two local minima of the potential $V_M(x_1,x_2)$. We sample a data set of $5000$ points for the state $A$ from the distribution: $\mathcal{U}\left(\mathcal{B}(a,r)\times[-2\sigma\sqrt{\epsilon},2\sigma\sqrt{\epsilon}]^8\right)$, where $\mathcal{U}(\cdot)$ denotes the uniform distribution and $\mathcal{B}(a,r)$ indicates a ball, $\mathcal{B}(a,r)= \{(x_1,x_2):|(x_1,x_2)- a|<r\}$. Similarly, we sample a data set for the state $B$. These two data sets will be used to impose the boundary conditions over $A$ and $B$ as in Eq.~\eqref{loss_bd} for computing the committor function. 

\begin{figure}[t!]
\centering
\includegraphics[width=\linewidth]{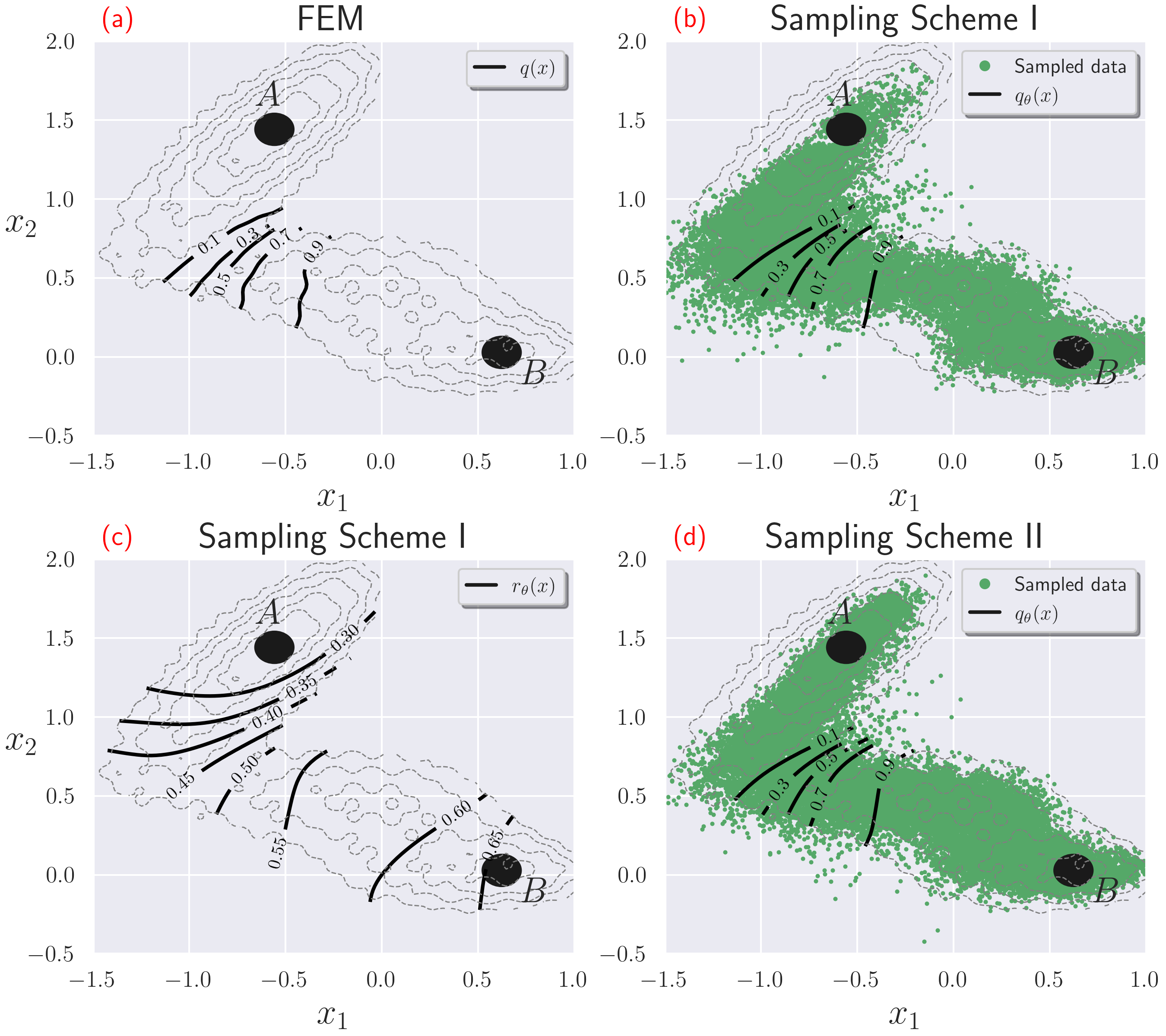}
\caption{(a) Contour plots of the committor function $q(x)$ computed using the finite element method. (b) Distribution of $5\times 10^4$ data points generated by sampling scheme I in \cref{alg1} and contour plots of the learned committor $q_{\theta}$ obtained using the sampled data. (c) Contour plots of the variable $r_{\theta}=R_{10}(q_{\theta})$ where $q_{\theta}$ is obtained using sampling scheme I. (d) Distribution of $5\times 10^4$ data points generated by sampling scheme II in \cref{alg2} and contour plots of the learned committor $q_{\theta}$ obtained using the sampled data. The data and functions are projected on the $(x_1,x_2)$-plane. The dashed contour lines indicate the potential energy~\eqref{example1} of the system.}
% \caption{Distribution of $5\times 10^4$ data points generated by sampling scheme I in \cref{alg1} ({\bf Middle}) or sampling scheme II in \cref{alg2} ({\bf Right}). Contour plots of the committor function $q(x)$ computed using the finite element method ({\bf Left}) and the learned committor $q_{\theta}$ obtained using the sampled data ({\bf Middle} and {\bf Right}). The data and functions are projected on the $(x_1,x_2)$-plane. The dashed contour lines indicate the potential energy~\eqref{example1} of the system.}
\label{fig1a}
\end{figure}

In the example, it can be shown that the committor function relies only on the first two dimensions, {\it i.e.} $q(x_1,\dots,x_{10})=\tilde{q}(x_1,x_2)$. One can compute a reference solution of $q$ using the finite element method~\cite{li2019computing}. Specifically, we solve a two-dimensional backward Kolmogorov equation associated with the potential $V_M(x_1,x_2)$ in a rectangle domain $\Omega_{12}=[-1.5,1]\times[-1,1.5]$ with Neumann boundary conditions. The finite element method is implemented using the software {\it FreeFem}~\cite{MR3043640}. 
A plot of the reference solution $q(x)$, projected on the $(x_1,x_2)$-plane, is shown in \cref{fig1a} (a). 
With $q$, one can evaluate the learned committor function $q_{\theta}$ using the relative error:
\begin{equation}\label{error}
    E(q_{\theta};\Omega') = \frac{\lVert q(x)-q_{\theta}(x)\rVert_{\Omega'}}{\lVert q(x)\rVert_{\Omega'}},
\end{equation}
where $\Omega'$ is a domain in the configuration space and $\lVert \cdot \rVert_{\Omega'}$ denotes the $L_2$ norm over $\Omega'$. In the example, we choose two different sets for $\Omega'$. First, we define a two-dimensional domain
\begin{equation*}
    \Omega'_{12}=\{(x_1,x_2)\in\Omega_{12}:|V_M(x_1,x_2)-\min_{\Omega_{12}} V_M(y_1,y_2)|\leq130\},
\end{equation*}
which excludes the points with potential far above from the minimum potential. The two sets are chosen as 
\begin{equation}\label{error_sets}
\Omega'_1=\Omega'_{12}\times[-2\sigma\sqrt{\epsilon},2\sigma\sqrt{\epsilon}]^8,\quad \Omega'_2=\{x\in\Omega'_1:|q(x)-0.5|\leq 0.2\}
\end{equation}
for computing the relative error of $q_{\theta}$ in Eq.~\eqref{error}. Note that $\Omega'_1$ contains most of the points in the transition tube of the system, while $\Omega'_2$ is chosen to focus on the $1/2$-isosurface of $q(x)$, {\it i.e.} the transition state region.

% We define a two-dimensional domain $\tilde{\Omega}_{12}=\{(x_1,x_2)\in\Omega_{12}:|V_M(x_1,x_2)-\min_{\Omega_{12}} V_M(y_1,y_2)|\leq130\}$, which excludes the points where the potential is far away from its minimum value. In what follows, we quantify the performance of the algorithm using two errors which are defined in Eq.~\eqref{error} over the two domains $\Omega_1=\tilde{\Omega}_{12}\times[-2\sigma\sqrt{\epsilon},2\sigma\sqrt{\epsilon}]^8$ and $\Omega_2=\{x\in\Omega_1:|q(x)-0.5|\leq 0.2\}$. The latter domain consists of the points locating in a neighborhood of the $1/2$-isocommittor surface of $q(x)$.

% \begin{figure}[t!]
% \centering
% \includegraphics[width=\linewidth]{Example1c}
% \caption{Plots of the relative errors of the learned committor function $q_{\theta^k}(x)$ versus the training step $k$ using \cref{alg1} with different numbers of $n$ ($n=1,1.5,2,5,10$) or \cref{alg2}. The errors as defined in Eq.~\eqref{error} are computed over the domain $\Omega_1$ ({\bf Left}) or the domain $\Omega_2$ ({\bf Right}).
% }
% \label{fig1c}
% \end{figure}

We apply \cref{alg1} and \cref{alg2} with $10$ iterations to compute the committor function at $\epsilon=10$. 
In \cref{alg1}, we generate data using sample scheme I. Specifically, we perform metadynamics with the variable $r_{\theta}=R_n(q_{\theta})$, $n=10$, where $2000$ Gaussian hills with height $h=2$ and width $w=0.003$ are deposited. The time step is set as $10^{-5}$ and the Gaussian hill is added for every $500$ steps. We sample $5\times 10^4$ data points from the equilibrium distribution~\eqref{rho_1} by generating trajectories following the dynamics with the modified potential and only keep those points with coordinates $x$ outside $A$ and $B$. With the sampled data, we train the neural network by minimizing the loss~\eqref{loss1}. In \cref{alg2}, we generate data using sample scheme II. First, we compute the free energy with $r_{\theta}$ by performing metadynamics with the same parameters as in \cref{alg1}. Then the free energy with $q_{\theta}$ and its derivative are obtained using Eq.~\eqref{r_q}. We sample $5\times 10^4$ data points from the equilibrium distribution~\eqref{rho_2} and train the neural network by minimizing the loss~\eqref{loss2}.

\begin{figure}[t!]
\centering
\includegraphics[width=\linewidth]{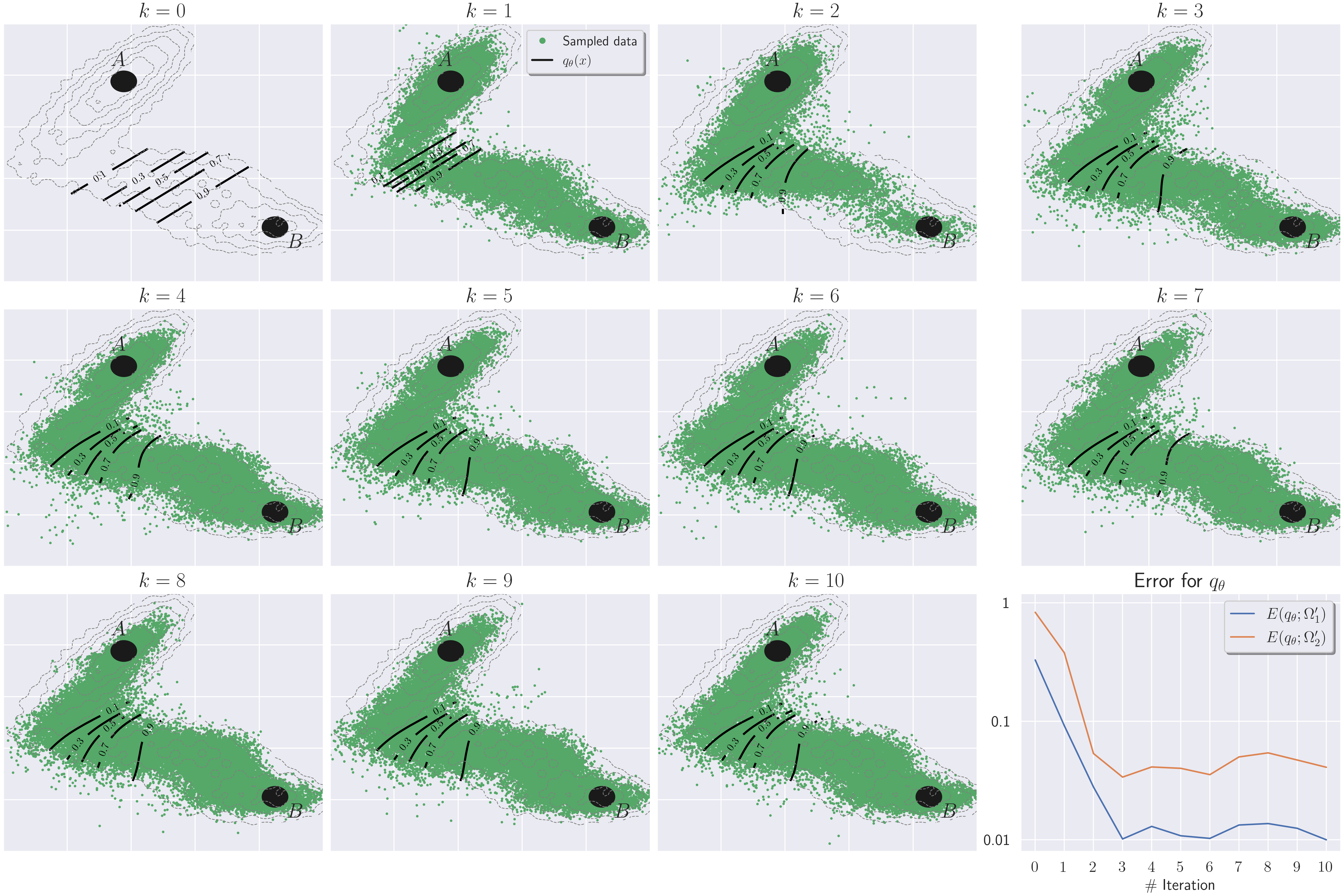}
\caption{Distribution of the sampled data points and contour plots of the learned committor $q_{\theta}$, projected on the $(x_1,x_2)$-plane, over the all iterations of \cref{alg2} with sampling scheme II. 
The first panel shows contour plots of the initial network for $q_{\theta}$. The last panel shows plots of the two errors for $q_{\theta}$ versus the iteration, which is defined by Eq.~\eqref{error} and \eqref{error_sets}.}
% \caption{Distribution of the sampled data points and contour plots of the learned committor $q_{\theta}$, projected on the $(x_1,x_2)$-plane, in the first $6$ iterations of \cref{alg2} with sampling scheme II. 
% The first panel shows contour plots of the initial network for $q_{\theta}$. The last panel shows plots of the two errors for $q_{\theta}$ versus the iteration, which is defined by Eq.~\eqref{error} and \eqref{error_sets}.}
\label{fig1b}
\end{figure}

In \cref{fig1a} (b) and (d), we show the distribution of the data, projected on the $(x_1,x_2)$-plane, obtained in the last iteration of the two algorithms. The results indicate that both sampling schemes are able to generate uniform data points in a tube connecting the two metastable states $A$ and $B$. This tube corresponds to the transition tube of the system. 
Also shown in the figure are the learned committor function $q_{\theta}(x)$ (panel (b) and (d)) for both sampling schemes and the corresponding variable $r_{\theta}(x)$ (panel (c)) for sampling scheme I. From the figure, one can observe that $q_{\theta}$ agrees well with the reference solution $q$ for both sampling schemes, and the iso-surfaces of $r_{\theta}$ corresponding to uniform target values are distributed almost uniformly on the $(x_1,x_2)$-plane. 
% uniformly in the configuration space as compared to those of $q_{\theta}$.
The latter enables one to sample uniform data along the transition tube using the metadynamics with $r_{\theta}$. 
The relative errors for $q_{\theta}$ over $\Omega'_1$, $\Omega'_2$ as defined by Eq.~\eqref{error} and \eqref{error_sets} are $0.0094$, $0.0337$ for sampling scheme I and $0.0100$, $0.0409$ for sampling scheme II, respectively. The results demonstrate the accuracy of the method for computing the committor function of the high-dimensional system. To illustrate the whole training process of the method, we show the distribution of sampled data and the learned committor $q_{\theta}$ over the all iterations of \cref{alg2} in \cref{fig1b}. We also plot the relative errors for $q_{\theta}$ versus the iteration in the last panel of \cref{fig1b}. 
The results show that after $3$ iterations, the data distribution, solution $q_{\theta}$ and errors vary in small magnitudes, 
% do not vary too much and the error is kept as small. 
which demonstrates the stability of the method. 

% Also shown in \cref{fig1c} are the errors of the learned committor function using \cref{alg1} with other values of $n$ ($n=1,1.5,2,5$). The results indicate that it is needed to choose a suitable $n\geq 2$ so that the algorithm is getting close to the ground truth solution for this example, although the accuracy of the algorithm may not vary too much for large $n$.

% \begin{figure}[t!]
% \centering
% \includegraphics[width=\linewidth]{fig1c}
% \caption{Distribution of $5\times 10^4$ data points generated by umbrella sampling ({\bf Left}), sampling scheme I ({\bf Middle}) or sampling scheme II ({\bf Right}) and contour plots of the learned committor $q_{\theta}$ obtained using the sampled data.}
% \label{fig1c}
% \end{figure}

We remark that, in principle, the proposed method is able to sample uniform data along the transition tube, whereas the previous umbrella sampling method may require a fine tuning of parameters to sample the entire transition tube~\cite{rotskoff2022active,hasyim2022supervised}. 
We quantitatively compare the performance of sampling scheme I and II and the umbrella sampling method with ten windows for computing the committor function. 
% To illustrate the advantages of sampling scheme I and II over the umbrella sampling method, we compute the error for $q_{\theta}$ obtained using the three sampling methods. 
Specifically, we first approximate the reference solution $q(x)$, which is obtained using the finite element method, by a neural network $q^0_{\theta}(x)$ using supervised learning. Then we generate $5\times 10^4$ data points using the three sampling methods, respectively, and train the neural network with the corresponding loss. The details for the supervised learning and umbrella sampling are introduced in \cref{App_B}. The relative error of $q_{\theta}$ over $\Omega'_1$ is $0.4750$ for the umbrella sampling, $0.0120$ for sampling scheme I and $0.0131$ for sampling scheme II, respectively. The results demonstrate the advantages of the current method over the previous adaptive sampling method for computing the committor function.

\subsection{Alanine Dipeptide}
To show the applicability of the method to molecular systems, we apply the method to study the isomerization of the alanine dipeptide in vacuum. 
% in vacuum at room temperature $T=300K$ using the two algorithms. 
This is a molecule which has been extensively used as a benchmark system in a number of works since it is a relatively simple while exhibiting common features to general bio-molecules~\cite{apostolakis1999calculation,bolhuis2000reaction,ma2005automatic,ren2005transition,maragliano2006string,evans2023computing}. 
Most of the previous works for investigating the transition mechanism 
% underlying the alanine dipeptide 
are based on the availability of prior knowledge regarding the transition~\cite{maragliano2006string,li2019computing,evans2023computing}, for instance, the collective variables for describing the transition which includes a few torsion angles of the molecule. In this work, the method does not require the prior knowledge. Instead, the nature of adaptive sampling and the two sampling schemes in the method enable us to generate uniform data along the transition tube by making use of the learned committor function, which will be illustrated below.

\begin{figure}[t!]
\centering
\includegraphics[width=.8\linewidth]{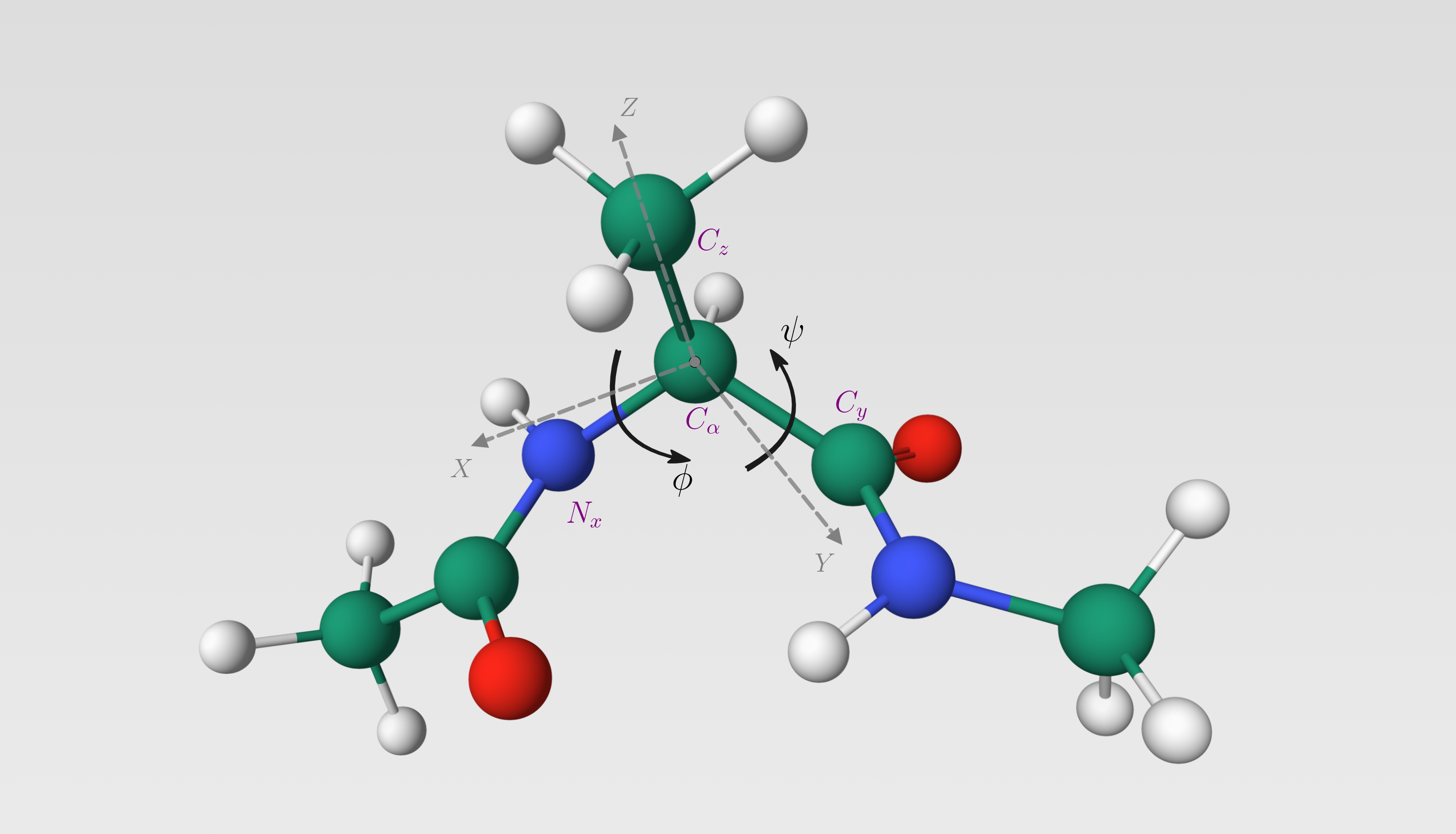}
\caption{Stick and ball plot of the alanine dipeptide (C$\text{H}_3$-CONH-CHC$\text{H}_3$-CONH-C$\text{H}_3$). In the neural network architecture of $q_{\theta}(x)$, the input configuration $x$ is transformed through translation and rotation with respect to the four atoms $C_{\alpha}$, $C_{z}$, $N_{x}$ and $C_{y}$, as described in Eq.~\eqref{ala_q}.}
% The four atoms labeled as $C_{\alpha}$, $C_{z}$, $N_{x}$ and $C_{y}$ are used in the neural network representation of the committor function.
\label{fig2a}
\end{figure}

The alanine dipeptide has $22$ atoms and a schematic plot of the molecule is shown in \cref{fig2a}.
% shows a stick and ball plot of the molecule. 
In vacuum, the system exhibits two typical metastable states $C_{7eq}$ and $C_{ax}$ at the room temperature $T=300$ K. 
% and there are two distinct transition pathways connecting the two states. 
To describe the two states, we denote by $C_{\alpha}$ the central carbon atom in the molecule and refer to the torsion angles corresponding to the quadruples of atoms containing $C_{\alpha}$, $(C,N,C_{\alpha},C)$ and $(N,C_{\alpha},C,N)$ as $\phi$ and $\psi$, respectively.
% . In the molecule, the two sets of atoms $(O,C,N,H)$ which are respectively located on the left and right side of the atom $C_{\alpha}$ are referred to as the two dipeptide groups of the molecule. In particular, the two torsion angles $\phi(x)$ and $\psi(x)$ are quadruples of the atoms: $(C,N,C_{\alpha},C)$ for $\phi$ and $(N,C_{\alpha},C,N)$ for $\psi$, as illustrated in \cref{fig2a}. 
Then, we define the metastable sets $A$ and $B$ as 
\begin{equation}
\begin{aligned}
    A &= \{x\in\R^{66}:|(\phi(x),\psi(x))-a|<r\},\\
    B &= \{x\in\R^{66}:|(\phi(x),\psi(x))-b|<r\},
\end{aligned}
\end{equation}
where $a=(-85^{\circ},75^{\circ})$, $b=(72^{\circ},-75^{\circ})$ and $r=10^{\circ}$.

% It is a general physical property for molecules that they are equivalent to translation and rotation of the configuration. 
In the example, the system is equivalent to any translation and rotation of the molecule in the configuration space.
To enhance the training efficiency, we make use of this physical property in establishment of the neural network for representing the committor function. Specifically, we first define a transformation function $T(x)$ which is a composite of translation and rotation of the molecule. For a given configuration $x$, the atoms $C_{\alpha}$, $C_z$, $N_x$ and $C_y$ in the new configuration $T(x)$, as depicted in \cref{fig2a}, are located in the origin, $Z$-axis, $XZ$-plane and the region where the $Y$-component is positive, respectively. 
With the given function $T(x)$, we parameterize the committor function as
\begin{equation}\label{ala_q}
    q_{\theta}(x) = F_{\theta}(T(x)),\quad x\in\R^{66}
\end{equation}
where $F_{\theta}$ is a fully-connected neural network with two hidden layers and $100$ nodes on each hidden layer. We put a sigmoid function on the output layer so that $q_{\theta}(x)\in[0,1]$. We next apply \cref{alg1} and \cref{alg2} with $K=10$ iterations to compute the committor function at the room temperature.

\begin{figure}[t!]
	\centering
	\includegraphics[width=\linewidth]{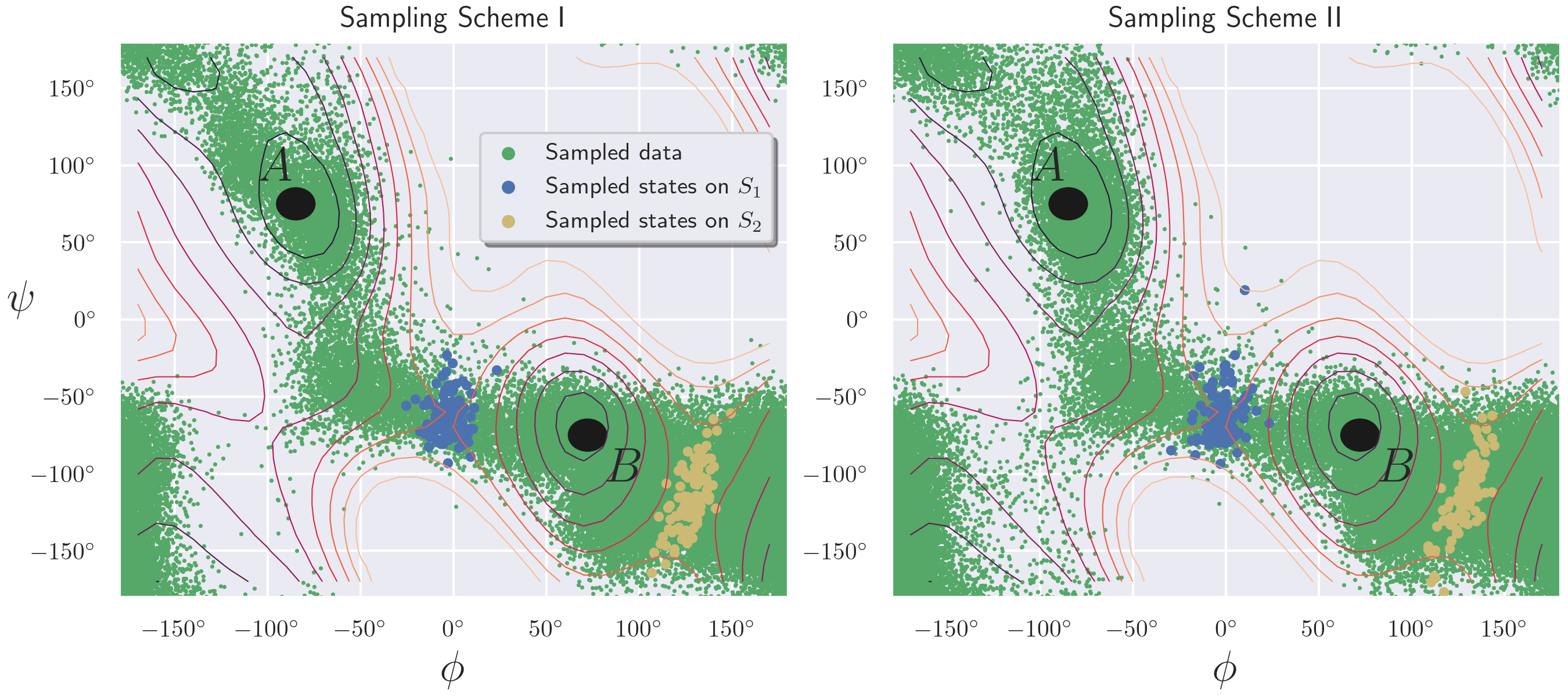}
	\caption{Distribution of $5\times 10^4$ data points generated using sampling scheme I ({\bf Left}) or II ({\bf Right}) in the algorithms and $200$ states sampled on the $1/2$-isosurface of $q_{\theta}$, projected on the $(\phi(x),\psi(x))$-plane. The isosurface consists of two separate sets $S_1$ and $S_2$, each set with $100$ states sampled. The contour lines indicate the adiabatic energy landscape of the alanine dipeptide which is computed by energy minimization with $\phi$ and $\psi$ fixed.}
	\label{fig2b}
\end{figure}

In the example, the molecular dynamics simulation is performed using the package {\it NAMD} \cite{phillips2005scalable} with the CHARMM22 all-hydrogen force fields. We set the damping coefficient in the Langevin dynamics as $1$/ps in the simulation. First, we sample $2000$ data points for each of the two metastable sets $A$ and $B$ by generating short trajectories following the dynamics starting from the set. The two data sets are used in the loss~\eqref{loss_bd} to impose the boundary conditions. 
In \cref{alg1}, we generate data using sampling scheme I. Specifically, we perform metadynamics with the variable $r_{\theta}(x)=R_n(q_{\theta}(x))$, $n=10$, where we set the time step as $1$ fs and deposit $2000$ Gaussian hills, one for every $1000$ steps, with the height $h=0.15$ kcal/mol and width $w=0.003$.
Subsequently, we sample $5\times 10^4$ data points with the modified potential and only keep those data points with configuration outside $A$ and $B$. The neural network is trained by minimizing the loss~\eqref{loss1}. In \cref{alg2}, we generate data using sampling scheme II. First, we compute the free energy with $r_{\theta}(x)$ using metadynamics with the same parameters as in \cref{alg1}. Then we obtain the free energy with $q_{\theta}$ using Eq.~\eqref{r_q}. We sample $5\times 10^4$ data points from the equilibrium distribution~\eqref{rho_2} and train the neural network using the loss~\eqref{loss2}. 
In \cref{fig2b}, we show the distribution of the data, projected on the $(\phi(x),\psi(x))$-plane,  obtained in the last iteration of the two algorithms. 
From the figure, for both sampling schemes, the sampled data form a tube containing the two metastable states $A$ and $B$; in fact, this tube agrees well with the transition tube computed using the finite-temperature string method in Ref.~\cite{ren2005transition}. The results indicate that the sampling schemes are able to produce uniform data along the transition tube.  
% one can observe that both sampling schemes are able to produce uniform data points between the two states $A$ and $B$.  

\begin{figure}[t!]
	\centering
	\includegraphics[width=\linewidth]{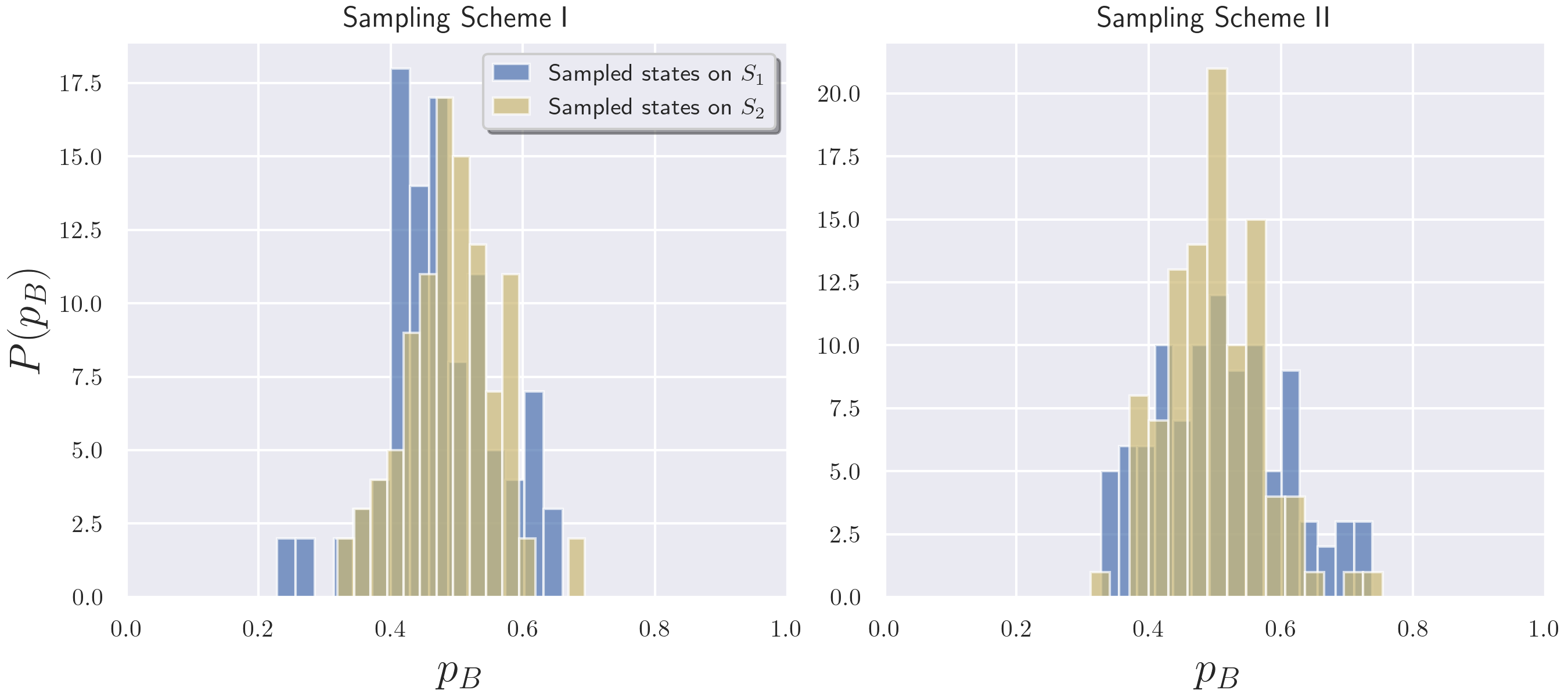}
	\caption{Distribution of the committor values for the $200$ states sampled on the $1/2$-isosurface of $q_{\theta}$ learned with sampling scheme I ({\bf Left}) or II ({\bf Right}). The committor values are computed using the Monte Carlo method.  The isosurface consists of two separate sets $S_1$ and $S_2$.}
 % {\color{red} change $C_{ax}$ to $p_B$}}
	\label{fig2c}
\end{figure}

For the molecular system, the analytic form or other numerical solution of the committor function for evaluating the learned committor $q_{\theta}$ is not straightforwardly available. To validate the solution $q_{\theta}$, we check whether the $1/2$-isosurface of $q_{\theta}$ locates the transition states~\cite{li2019computing}. Of the sampled data points as shown in \cref{fig2b}, there are two separate clusters on the $(\phi,\psi)$-plane where $q_{\theta}$ is close to $1/2$, indicating that the $1/2$-isosurface consists of two separate sets, denoted by $S_1$ and $S_2$. The two sets indeed correspond to two separate transition pathways of the system. 
% We pick up two states from the separate clusters, respectively. 
To sample states on the $1/2$-isosurface of $q_{\theta}$, we first pick up two states on $S_1$, $S_2$ and then perform the constrained dynamics simulation twice starting from the picked states, where an external function $V_q(x)=\frac{1}{2}\kappa (q_{\theta}(x)-\frac{1}{2})^2$ is added into the potential. The parameter $\kappa=10^3$ kcal/mol. 
For each simulation, we collect $100$ states from the generated trajectory. \cref{fig2b} shows the distribution of the $200$ sampled states on the $1/2$-isosurface of $q_{\theta}$. 
Then we compute the committor values of the $200$ states using the Monte Carlo method, {\it i.e.} by directly simulating the Langevin dynamics, according to Eq.~\eqref{def_q}. Specifically, we generate $500$ trajectories starting from each of the states with random initial velocities and compute the probability of the system first arriving in $B$ rather than $A$. 
In \cref{fig2c}, we plot a histogram of the computed committor values. The results show that the committor values cluster around $1/2$ on the two sets $S_1$ and $S_2$ for both sampling schemes used in the algorithms, which demonstrates the accuracy of the method for producing transition states of the molecular system.

\subsection{Solvated Dimer System}
% In this section, we apply the proposed method to a dimer system which is solvated in water.
% In this section, we compute the committor function for studying the transition between the two stable states of a dimer system in a solvent of purely repulsive particles.
To further illustrate the ability of the method for dealing with solvated systems, we apply the method to study the isomerization of a dimer immersed in a solvent of purely repulsive particles~\cite{dellago1999calculation,hasyim2022supervised}. 

In the example, we put the dimer in a three-dimensional cube with periodic boundary conditions. There are $N_a=30$ solvent particles in the cube. The length of the cube is set as $\mathcal{L}=10$ on each dimension. In the system, the particles interact via two types of bond-based potentials. Specifically, the two dimer particles interact via the double-well potential
\begin{equation*}
    V_{\text{DW}}(d) = a\left[1-\frac{(d-d_{\text{WCA}}-s_{\text{DW}})^2}{s_{\text{DW}}^2}\right]^2,
\end{equation*}
where $d$ denotes the distance between the two particles in the water box, $d_{\text{WCA}}=2^{1/6}$, $s_{\text{DW}}=0.25$ and $a=5$ specifies the energy height between the two local minima of the potential, $d=d_{\text{WCA}}$ and $d=d_{\text{WCA}}+2s_{\text{DW}}$. 
The interactions between the solvent particles and between the solvent and dimer particles are both described by the repulsive Weeks-Chandler-Anderson potential
\begin{equation*}
    V_{\text{WCA}}(d) = 
    \begin{cases}
        4b\left(\dfrac{1}{d^{12}}-\dfrac{1}{d^6}\right)+b, & d\leq d_{\text{WCA}}\\
        0,& d> d_{\text{WCA}}
    \end{cases}
\end{equation*}
where $b=1$ specifies the strength of the potential.

\begin{figure}[t!]
\centering
\includegraphics[width=.8\linewidth]{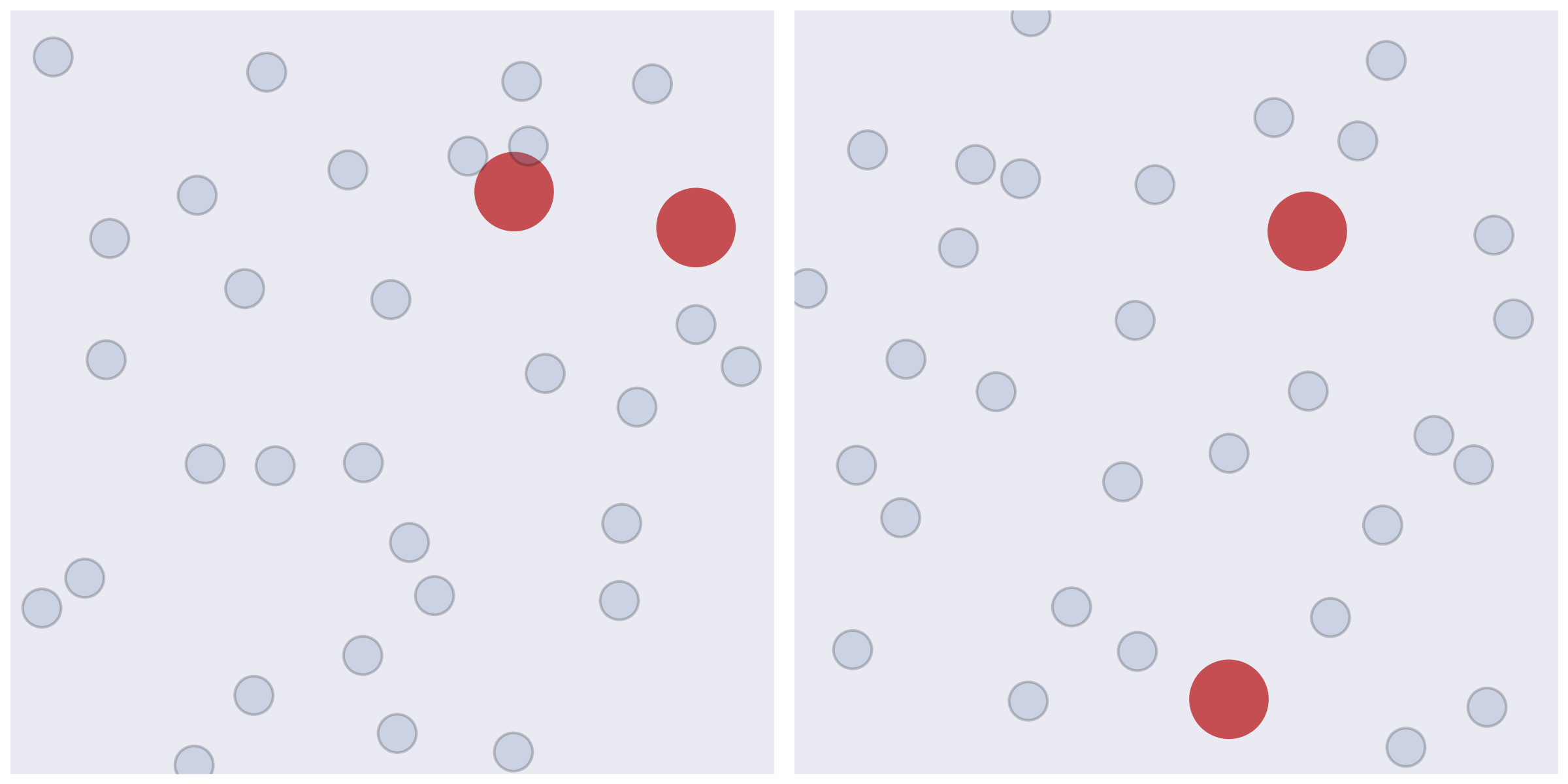}
\caption{Two typical configurations for the compacted (\textbf{Left}) and extended states (\textbf{right}) of the dimer system immersed in a cube with solvent particles, where all particles are projected on a two-dimensional plane. The big red circles and small transparent circles represent the dimer particles and solvent particles, respectively.}
\label{fig3a}
\end{figure}

The system has two metastable states, referred to as the compacted and extended states~\cite{dellago1999calculation}. We defined the two states as $A=\{x\in \R^{96}:\mu(x)\leq d_{\text{WCA}}\}$ and $B=\{x\in \R^{96}:\mu(x)\geq d_{\text{WCA}}+2s_{\text{DW}}\}$, where $\mu(x)$ is the bond length between the two dimer particles. \cref{fig3a} shows two typical configurations corresponding to the two states, which are projected on a two-dimensional plane. Next, we apply \cref{alg1} and \cref{alg2} to compute the committor function between $A$ and $B$ at the temperature $\epsilon=1$. 

In the example, we set the neural network for approximating the committor function as Schnet~\cite{schutt2017schnet}. This network reflects the physical properties including invariance to rotation and translation of the system, as well as particle re-indexing, thus enabling enhancement of the training efficiency.  
In Schnet, all the particles are represented by a tuple of features through the layers. First, by an embedding based on particle types, the representation is initialized using two trainable 64-dimensional feature vectors for the dimer and solvent types, respectively. Then the feature representation is updated via two interaction blocks which models the interaction between particles, followed by two dense layers. The final output is obtained by a pooling of the resulting particle-wise features. 
In each interaction block, we take the following $100$ radial basis functions in expanding the distances between particles for the continuous-filter convolution layer: 
\begin{equation}\label{Expand_dist}
e_k(x_i,x_j)=\exp\left(-\frac{(|x_i-x_j|-c_k)^2}{\nu^2}\right),\quad 0\leq k\leq 99
\end{equation}
where the centers and width of the basis functions are set as $c_k=k\mathcal{L}'/99$, $\nu=\sqrt{10}\mathcal{L}'/99$ with $\mathcal{L}'=\sqrt{3}\mathcal{L}/2+1$, respectively. Here $\mathcal{L}'$ is taken to be greater than the maximal distance in the cube with length $\mathcal{L}$ and periodic boundary conditions and $\{c_k\}$ corresponds to the mesh points in a discretization of $[0,\mathcal{L}']$. Moreover, the expanded distances are transformed into filter weights through two dense layers. 
In the neural network, the number of feature maps are kept as $64$ in the interaction blocks and set to be $32$ and $1$ on the two subsequent dense layers. We use the hyperbolic tangent ($\tanh$) as the activation function and put a sigmoid function on the output layer.

\begin{figure}[t!]
\centering
\includegraphics[width=.8\linewidth]{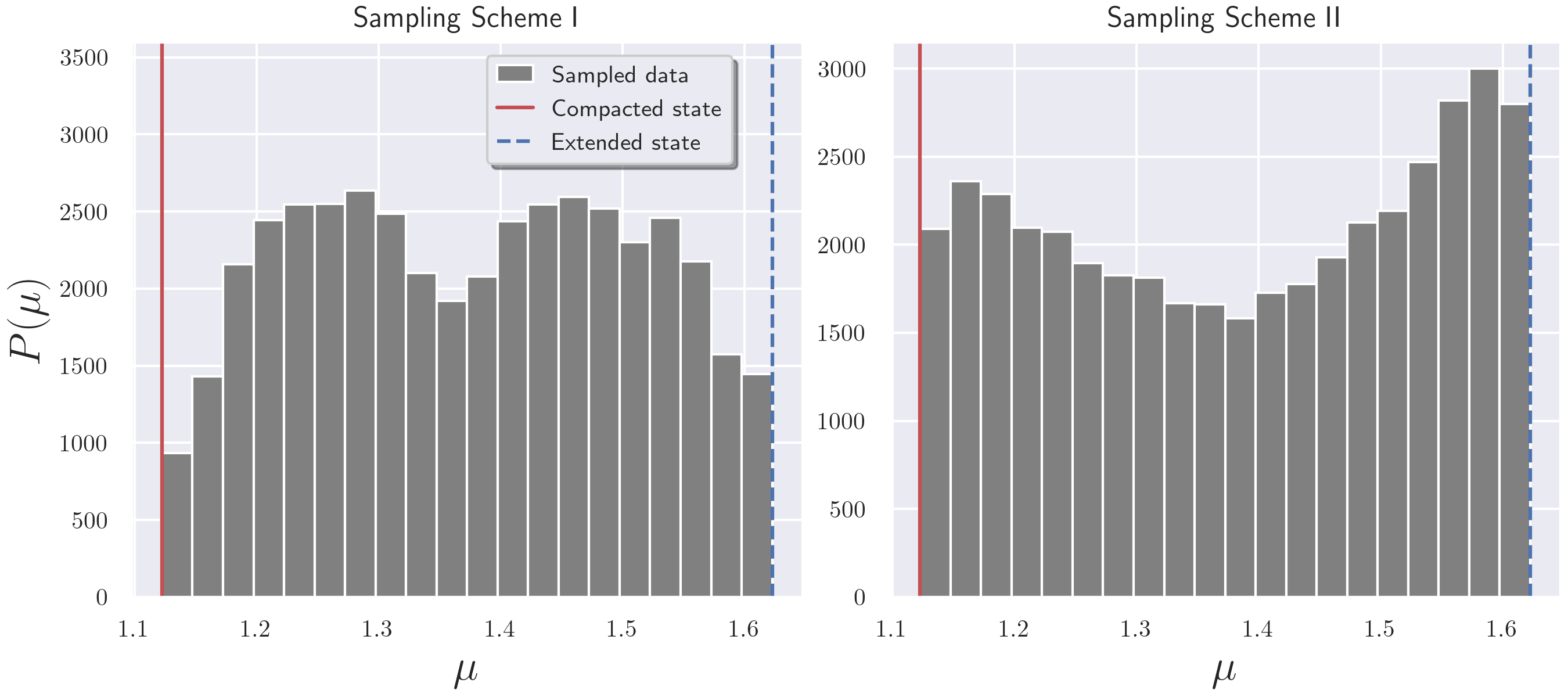}
\caption{Distribution of the bond length $\mu(x)$ (between the dimer particles) of $5\times 10^4$ data points generated using sampling scheme I ({\bf Left}) or II ({\bf Right}) in the algorithms. The solid red and dashed blue vertical lines indicate the bond lengths $d_{\text{WCA}}$ and $d_{\text{WCA}}+2s_{\text{DW}}$, respectively, for characterizing the compacted and extended states of the solvated dimer system.
}
\label{fig3b}
\end{figure}

\begin{figure}[t!]
\centering
\includegraphics[width=.8\linewidth]{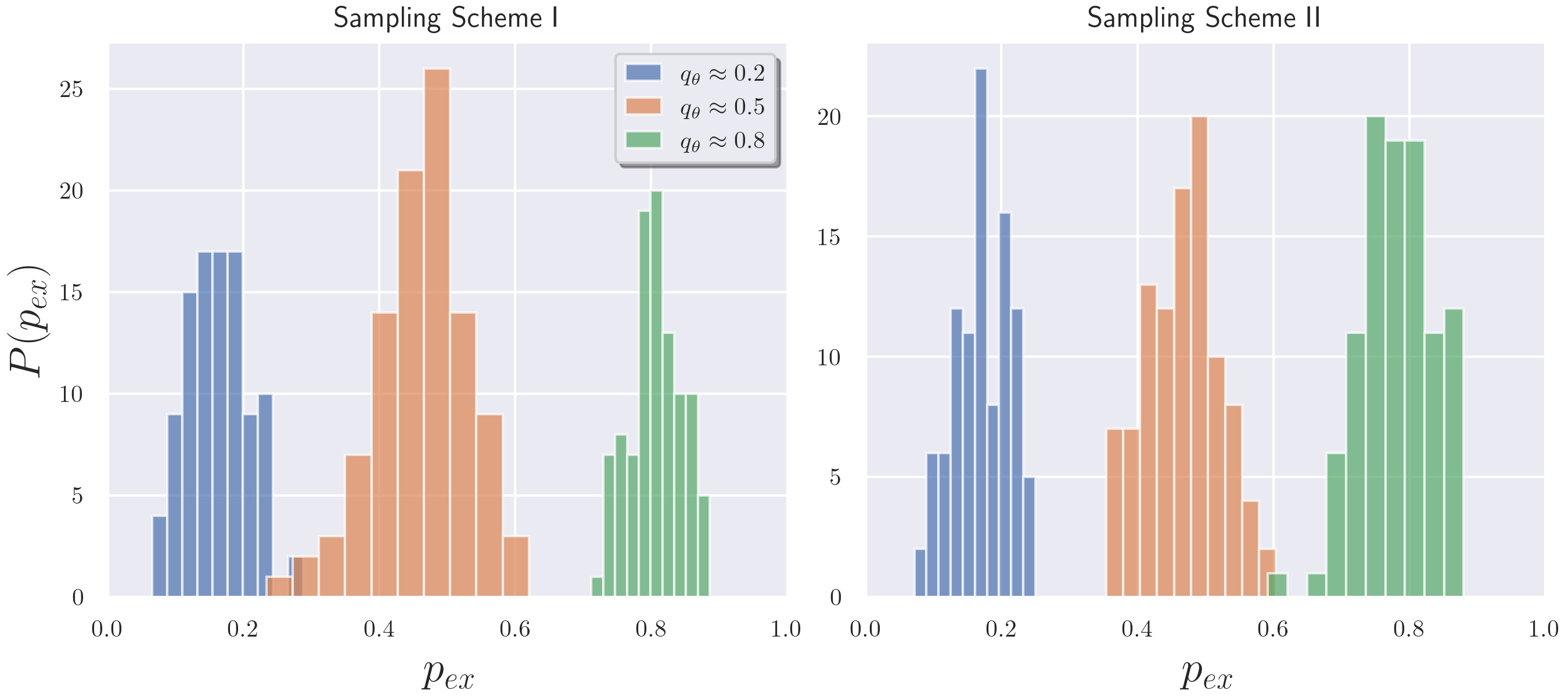}
\caption{Distribution of the committor values for the states sampled on the three isosurfaces of $q_{\theta}$ ($q_{\theta}=0.2, 0.5, \text{or } 0.8$) learned with sampling scheme I ({\bf Left}) or II ({\bf Right}). There are $100$ states sampled on each of the isosurfaces. The committor values are computed using the Monte Carlo method.}
\label{fig3c}
\end{figure}

For each of two metastable sets $A$ and $B$, we sample $2000$ data points in the set by generating short trajectories following the dynamics starting from this set at the temperature $\epsilon'=0.5$. 
% We sample two data sets $\mathcal{D}_A$, $\mathcal{D}_B$ in $A$, $B$, each set with $2000$ data points, by generating two short trajectories following the dynamics  starting from $A$ and $B$, respectively, at the temperature $\epsilon'=0.5$. 
We carry out $K=10$ iterations in \cref{alg1} and \cref{alg2} for computing the committor function. 
In \cref{alg1}, we generate data using sampling scheme I. Specifically, we perform metadynamics with the variable $r_{\theta}(x)=R_n(q_{\theta}(x))$, $n=10$, where we set the time step as $10^{-4}$ and deposit $1000$ Gaussian hills, one for every $200$ steps, with the height $h=0.1$ and width $w=0.003$. Then we sample $5\times 10^4$ data points with the modified potential and train the neural network using the loss~\eqref{loss1}. In \cref{alg2}, we generate data using sampling scheme II. We obtain the free energy with $q_{\theta}$ by first computing the free energy with $r_{\theta}(x)$ using metadynamics and Eq.~\eqref{r_q}. Then we sample $5\times 10^4$ data points from the equilibrium distribution~\eqref{rho_2} and train the neural network using the loss~\eqref{loss2}. To visualize the sampled data, we plot histograms of the bond length $\mu(x)$ of the data obtained in the last iteration of the two algorithms, respectively. From the figure, one can observe that both sampling schemes are able to produce uniform data points in a region connecting the compacted and extended states. 

We evaluate the solution $q_{\theta}$ by validating its accuracy on the three isocommittor surfaces where $q_{\theta}=0.2, 0.5, \text{or } 0.8$. As we did in the previous example, we sample $100$ states on each of the three isosurfaces by performing the constrained dynamics simulation with an additional function $V_q(x)=\frac{1}{2}\kappa (q_{\theta}(x)-\hat{q})^2$ in the potential, where the parameter $\kappa=10^3$ and $\hat{q}$ is the target value corresponding to the isosurface, {\it i.e.} $\hat{q}=0.2, 0.5, \text{or } 0.8$. Then we compute the committor values of the sampled states using the Monte Carlo method. \cref{fig3c} shows the distribution of the committor values, from which one can observe the values cluster around the corresponding target values ($0.2$, $0.3$ and $0.8$) for the three isocommittor surfaces. This demonstrates the accuracy of the method for computing the committor function of solvated systems.

\section{Conclusions}\label{conclusion}

In this work, we developed a deep learning method with two adaptive sampling schemes for computing the committor function. 
The motivation of the work is that sampling adequate data for transition is crucial for understanding the transition mechanism but a challenging task for complex systems. 
We theoretically showed the advantages of the sampling schemes and proved that the data in sampling scheme II are uniform along the transition tube. The distribution of the data in the two schemes and the accuracy of the method were also numerically demonstrated in high-dimensional systems including the alanine dipeptide and a solvated dimer system. 
The implementation of the sampling schemes is relatively simple, as it mainly involves metadynamics with a one-dimensional variable depending on the learned committor function

The nature of deep learning and adaptive sampling makes the method a promising way for exploring the transition tube and generating adequate samples along the tube, thus studying transitions in high-dimensional and complex systems. In the future, we intend to apply the method to more complex systems, such as studying the protein folding.

\section*{Acknowledgement}
The work of B. Lin and W. Ren is partially supported by A*STAR under its AME Programmatic programme: Explainable Physics-based AI for Engineering Modelling \& Design (ePAI) [Award No. A20H5b0142]. 

\bibliographystyle{plain}
\bibliography{refs}

\newpage
\appendix

\section{Proof of Eq.~\eqref{g1} and Eq.~\eqref{g2}}\label{App_A}
\subsection{Proof of Eq.~\eqref{g1}}
Let $\mathfrak{X}$ be a random variable with the probability distribution $p_{\text{I}}(x)$ in Eq.~\eqref{p1}. 
As $F_r$ is the free energy associated with the variable $r(x)$, the probability density function (PDF) of the variable $r(\mathfrak{X})$ is given by
\begin{equation*}
\begin{aligned}
    g_r(z) &\sim \int_{\Rd} \exp\left(-\frac{1}{\epsilon}\Big[V(x)-F_r(r(x))\Big]\right)\delta(r(x)-z) dx\\
    &=\exp\left(\frac{1}{\epsilon}F_r(z)\right)\int_{\Rd} \exp\left(-\frac{1}{\epsilon}V(x)\right)\delta(r(x)-z) dx\\
    &=1,
\end{aligned}
\end{equation*}
for $z\in [0,1]$. By change of variable, the PDF of the tube variable $s(\mathfrak{X}) = f^{-1}\circ q(\mathfrak{X})
= f^{-1}\circ R_n^{-1}(r(\mathfrak{X}))$ is given by
\begin{equation*}
\begin{aligned}
     g_{\text{I}}(z) &= g_r(z') \cdot (R_n\circ f)'(z)\\
          &= R_n'(f(z))f'(z)\\
          &\sim R_n'(f(z)) \exp(F(z)/\epsilon),\qquad z\in [0,1]
\end{aligned}
\end{equation*}
where $z'=R_n(f(z))$. The last equality is due to Eq.~\eqref{f_alp}.

\subsection{Proof of Eq.~\eqref{g2}}
Denote by $F_s$ the free energy associated with the tube variable $s(x)$. Recall that $F_q$ is the free energy with the committor function $q(x)$. First, we show that 
\begin{equation}\label{B1}
\begin{aligned}
    F_s(z) \approx F_q(f(z))/2,\quad z\in [0,1].
\end{aligned}
\end{equation}
Let $\mathfrak{Y}$ be a random variable with the equilibrium distribution $\rho(x)$ of the system. Since $s(\mathfrak{Y})=f^{-1}\circ q(\mathfrak{Y})$, by change of variable, we have
\begin{equation*}
\begin{aligned}
    \exp\left(-\frac{1}{\epsilon}F_s(z)\right)
    &\sim \exp\left(-\frac{1}{\epsilon}F_q(f(z))\right)
    f'(z) \\
    &\sim \exp\left(-\frac{1}{\epsilon}F_q(f(z))\right) \exp(F(z)/\epsilon)
\end{aligned}
\end{equation*}
Here $\exp\left(-F_s(\cdot)/\epsilon\right)$ and $\exp\left(-F_q(\cdot)/\epsilon\right)$ represent the PDF of the variable $s(\mathfrak{Y})$ and $q(\mathfrak{Y})$, respectively. 
From the above equation, one can deduce that
\begin{equation*}
\begin{aligned}
    F_s(z) = F_q(f(z)) - F(z).
\end{aligned}
\end{equation*}
Recall that $F$, as defined under Eq.~\eqref{f_alp}, is the free energy with the variable $\tilde{s}(x)=f^{-1}\circ \tilde{q}(x)$, which is close to $s(x)$. This implies $F$ is close to the free energy $F_s$. Therefore, Eq.~\eqref{B1} is proved.

Now let $\mathfrak{Z}$ be a random variable with the probability distribution~\eqref{p2}. The PDF of the tube variable $s(\mathfrak{Z})$ can be written as
\begin{equation*}
\begin{aligned}
    g_{\text{II}}(z) &\sim \int_{\Rd} \exp\left(-\frac{1}{\epsilon}\Big[V(x)-\frac{1}{2}F_q(q(x))\Big]\right)\delta(s(x)-z) dx\\
    &=\exp\left(\frac{1}{2\epsilon}F_q(z)\right)\int_{\Rd} \exp\left(-\frac{1}{\epsilon}V(x)\right)\delta(s(x)-z) dx\\
    &\approx \exp\left(\frac{1}{\epsilon}F_s(z)\right)\int_{\Rd} \exp\left(-\frac{1}{\epsilon}V(x)\right)\delta(s(x)-z) dx\\
    &=1,
\end{aligned}
\end{equation*}
for $z\in [0,1]$. This shows that data in sampling scheme II are uniform along the transition tube.

\section{Details of the supervised learning and umbrella sampling for the Mueller system in Section~\ref{Mueller}}\label{App_B}
\subsection{Supervised Learning} 
We approximate the committor function by a neural network $q^0_{\theta}(x)$ using supervised learning. 
% The committor function is parameterized by a neural network $q_{\theta}(x)$ with architecture as described in the paper. 
To this end, we sample a set of $M=10^5$ data points, $\{X_i\}_{1\leq i \leq M}$ from the uniform distribution over the domain $\Omega_1'$ defined in Eq.~\eqref{error_sets}. 
Denote by $\{q_i\}_{1\leq i \leq M}$ the reference solution $q$ for the committor function on these points which is 
computed using the finite element method as described in the paper, {\it i.e.} $q_i=q(X_i)$.
With the data set $\{X_i,q_i\}_{1\leq i \leq M}$, we train the neural work $q^0_{\theta}$ using the regression loss
\begin{equation*}
    L_{SL}(\theta) = \frac{1}{M}\sum_{i=1}^M |q^0_{\theta}(X_i)-q_i|^2.
\end{equation*}
After $5\times 10^4$ training steps, the loss is small and on the order of $10^{-6}$. We obtain the approximated committor function $q^0_{\theta}(x)$. 

\subsection{Umbrella sampling}
With $q^0_{\theta}$, we generate data using the umbrella sampling method~\cite{rotskoff2022active,hasyim2022supervised} with $L=10$ windows. 
On each window, we sample a set of $N_l=5000$ data points from the distribution
\begin{equation*}
    p_l(x) \sim \rho(x) \exp\left(-\frac{1}{\epsilon}\kappa(q^0_{\theta}(x)-q_l)^2\right),\quad 1\leq l\leq L
\end{equation*}
with $\kappa=3000$ and the target value $q_l=(l-1)/9$, $1\leq l\leq L$. This is done by generating trajectories following the dynamics 
\begin{equation*}
    d x = -\nabla V_l(x) dt + \sqrt{2 \epsilon} d W_t,\quad t>0
\end{equation*}
and collecting states along the trajectories, where the modified potential $V_l(x) = V(x) +2\kappa(q^0_{\theta}(x)-q_l)^2$. 
Denote by $\mathcal{D}_l=\{X_l^j\}_{j=1}^{N_l}$ the data sampled for the window $l$. In total, there are $5\times 10^4$ points in the whole data set. To further train the neural network, we take the loss function from the functional~\eqref{prob} by reweighting the samples,
\begin{equation}\label{loss_app}
    \int_{\Omega\setminus(A\cup B)}  \lvert \nabla q_{\theta}(x) \rvert ^2 \rho(x) dx \approx \frac{\sum_{l=1}^L \frac{z_l}{N_l}\sum_{j=1}^{N_l} 
    \frac{\lvert \nabla q_{\theta}(X_l^j) \rvert ^2}{c(X_l^j)} }
    {\sum_{l=1}^L \frac{z_l}{N_l}\sum_{j=1}^{N_l} 
    \frac{1}{c(X_l^j)}},
\end{equation}
where the function $c(x)=\sum_{l=1}^L p_l(x)$ and the coefficients $\{z_l\}_{l=1}^L$ can be computed by solving the linear system
\begin{equation*}
    z_l = \sum_{l'=1}^L F_{l'l} z_{l'},\quad 1\leq l\leq L
\end{equation*}
with the condition $\sum_{l=1}^L z_l = 1$, where $F_{l'l}=\frac{1}{N_{l'}}\sum_{j=1}^{N_{l'}}p_{l}(X_{l'}^j)/c(X_{l'}^j)$. We train the neural network starting from $q^0_{\theta}$ by minimizing the loss in Eq.~\eqref{loss_app} for $2\times 10^4$ steps.

\end{document}